\documentclass[11pt]{article}

\usepackage{jheppub}
\usepackage{amsmath}
\usepackage{bm}
\usepackage{slashed}

\newcommand{\tr}{\mathrm{tr}\,}
\newcommand{\Tr}{\mathrm{Tr}\,}
\newlength{\dummysp}
\newcommand{\beq}{\begin{eqnarray}}
\newcommand{\eeq}{\end{eqnarray}}

\newcommand{\gappeq}{\mathrel{\rlap {\raise.5ex\hbox{$>$}}
{\lower.5ex\hbox{$\sim$}}}}
\newcommand{\lappeq}{\mathrel{\rlap{\raise.5ex\hbox{$<$}}
{\lower.5ex\hbox{$\sim$}}}}

\newcommand{\ben}{\begin{enumerate}}
\newcommand{\een}{\end{enumerate}}

\newcommand{\bit}{\begin{itemize}}
\newcommand{\eit}{\end{itemize}}

\newcommand{\avg}[1]{\left\langle#1\right\rangle}
 
\def\[{\left [}
\def\]{\right ]}
\def\({\left (}
\def\){\right )}
\def\R{{\mathbb R}}
\def\S{{\mathbb S}}
\def\Z{{\mathbb Z}}
\pdfoutput=1

\title{(S)QCD on $\mathbf{\R^3 \times \S^1}$: Screening of Polyakov loop by fundamental quarks and the demise of semi-classics}

\author[a]{Erich Poppitz,} \author[b]{Tin Sulejmanpasic}

\affiliation[a]{Department of Physics,   University of Toronto, 
Toronto, ON M5S 1A7, Canada}
\affiliation[b]{Institute for Theoretical Physics, University of
  Regensburg, 93040 Regensburg, Germany}
\emailAdd{poppitz@physics.utoronto.ca}\emailAdd{tin.sulejmanpasic@physik.uni-regensburg.de}

\abstract
{Recently, it was argued that the thermal deconfinement transition in pure Yang-Mills  theory is continuously connected to a quantum phase transition in softly-broken ${\cal{N}}$$=$$1$ supersymmetric pure YM  theory on $\R^{3} \times \S^1$. The transition  is semiclassically calculable at small  $\S^1$ size $L$,  occurs as the soft mass $m_{soft}$  and $L$ vary, and is driven by a competition between perturbative effects and nonperturbative  topological  molecules. These are correlated instanton-antiinstanton tunneling events whose constituents  are  monopole-instantons ``bound''  by attractive long-range forces. The mechanism driving the transition is universal for all simple gauge groups, with or without a center, such as $SU(N_c)$ or $G_2$. Here, we consider theories with fundamental quarks. We  examine the role  topological objects play in determining the fate of the (exact or approximate) center-symmetry in    $SU(2)$ supersymmetric  QCD (SQCD) with fundamental flavors, with or without soft-breaking terms.  In theories whose large-$m_{soft}$ limit is  thermal nonsupersymmetric QCD with massive quarks, we find a crossover of the Polyakov loop, from approximately center-symmetric at small ${1\over L}$   to maximally center-broken at larger $1\over L$, as seen in  lattice thermal QCD with massive dynamical quarks and $T={1 \over L}$.  We  argue that in all calculable cases, including  SQCD with exact center symmetry, quarks deform instanton-monopoles by  their  quantum fluctuations and do not   contribute to their binding. The semiclassical approximation and the molecular picture of the vacuum fail, upon decreasing the quark mass,  precisely when   quarks  would begin  mediating a long-range attractive force between monopole-instantons,  calling for a dual description of the resulting strong-coupling theory.}
  
\begin{document}
\maketitle
\flushbottom

\section{Introduction}

The electric deconfined phase of QCD and pure Yang-Mills theory at high temperatures has been understood as a result of the perturbative gluon screening potential \cite{Gross:1980br} for the Polyakov loop,  favoring a center-broken (i.e., deconfined) phase. The much more interesting confined phase, although studied extensively in lattice simulations, remains elusive due to the strong coupling and unreliable perturbative expansion. Further, the appearance of magnetic confinement, spatial area law, and magnetic mass gap  in the deconfined phase is also not analytically tractable even at high temperature.\footnote{Studies of the Polyakov loop potential have employed lattice gauge theory, field theoretical models, and  functional renormalization group (an incomplete list for thermal $\R^3 \times \S^1$ is \cite{Pisarski:2001pe, Fukushima:2003fw, Ratti:2005jh, Diakonov:2012dx, Greensite:2012dy, Haas:2013qwp}), while in thermal theories on $\S^3 \times \S^1$ or $\R^2 \times \S^1 \times \S^1$, it can be studied semi-classically (here, we  only quote the most recent studies of these two cases, \cite{Aharony:2006rf} and \cite{Anber:2012ig}, respectively, and refer the reader to the literature cited there).}

Supersymmetric theories on $\R^3\times \S^1$, on the other hand, are much more tractable by analytic tools, due to the ``power of holomorphy", allowing the study of many features of their infrared dynamics. A further simplification occurs 
at small $\S^1$-size $L$, where weak coupling methods  often  apply. There, the perturbative screening of the Polyakov loop is cancelled by the gluinos (it must be stressed that in order for this cancellation to be exact, a non-thermal, periodic compactification of the gluinos is necessary). Once the perturbative contribution is under control, the non-perturbative, exponentially suppressed instanton-monopole contribution becomes dominant. However, due to the presence of gaugino zero-modes, the instanton-monopoles cannot  directly generate  the bosonic potential for the Polyakov loop.

It was shown in \cite{Davies:1999uw} by an explicit calculation, following  \cite{Seiberg:1996nz,Aharony:1997bx}, that the resummation of instanton-monopole contributions gives rise to a bosonic potential for the gauge field holonomy around $\S^1$ and for the dual photon. Center-symmetry was not discussed there 
and the form of the bosonic potential was obtained from the  superpotential, thus relying entirely on supersymmetry. While giving the correct form for the potential, the underlying physical mechanism leading to center-symmetry stabilization and mass for the dual photon (i.e., confinement) was not discussed. The mechanisms responsible for these two phenomena were identified later and were argued to transcend supersymmetry. The dual photon (or magnetic) mass arises from the formation of the so-called ``magnetic bions", a kind of monopole-antimonopole molecules  \cite{Unsal:2007jx}.  The magnetic bions  are responsible for confinement in spatially compactified Yang-Mills theories with adjoint matter,  generalizing the 3d Polyakov confinement mechanism \cite{Polyakov:1976fu} to locally 4d theories. 
The mechanism for center-symmetry stabilization was, following \cite{Poppitz:2011wy}, elucidated in \cite{Poppitz:2012sw,Poppitz:2012nz}, as arising from the formation of another kind of monopole-antimonopole molecules called ``neutral bions", or ``center-stabilizing bions".\footnote{For conjectures on their role, including 2d and 4d theories without  supersymmetry, see \cite{Argyres:2012vv, Argyres:2012ka,Dunne:2012ae,Dunne:2012zk}.}  

Once the SYM theory is perturbed by a small gaugino mass $m_{soft}$ (which now controls the strength of the perturbative screening potential for the Polyakov loop), upon increasing $m_{soft}$ (at fixed $L$), center-symmetry gets destabilized due to both monopole-instanton and perturbative contributions to the potential. This gives rise to a center-symmetry breaking phase transition, as discussed in detail in \cite{Poppitz:2012sw,Poppitz:2012nz}, following  \cite{Unsal:2010qh}. It was conjectured there, and evidence was presented, that this center-symmetry breaking transition is continuously connected to the thermal deconfinement transition in pure Yang Mills theory, upon decoupling the gaugino. These results are reviewed in Section \ref{symreview}.

Our purpose here is to extend these studies to the case with fundamental matter. This is closer to the real-world QCD compared to studies previously done and may shed some light on the underlying mechanism of confinement/deconfinement in QCD.

\section{Summary and outline}

This paper is organized as follows. We first review the phase diagram of softly-broken SYM on $\R^3 \times \S^1$ in Section \ref{symreview}, along with the available evidence for a continuous connection to deconfinement in pure YM theory. In Section \ref{towards}, we begin  by reviewing the matter content of $SU(2)$ SQCD with $N_f$ flavors, the various boundary conditions on the matter fields that we consider, and the fields relevant for the description of the long-distance dynamics on $\R^3 \times \S^1$. The (s)quark contribution to the perturbative potential for the Polyakov loop, which arises upon introducing soft-breaking is computed in Appendix \ref{taming}.  In Section \ref{fundsymm}, we discuss the symmetries and the zero-modes of  instanton-monopoles. In Section \ref{qualitative}, without yet worrying about the constraints of supersymmetry, we consider the qualitative expectation for the center-symmetry realization in massive SQCD, following the logic of the earlier discussion of SYM theory. 

The main results of this paper are contained in Section \ref{massive}. We begin by  recalling some known exact results in SQCD on $\R^3\times \S^1$ (the ones that we will need) and set the stage for the consistent semiclassical computation of the superpotential of massive SQCD. In Section \ref{susyvacuum}, we give the result of this computation and explain the meaning of all terms. The somewhat lengthy technical details containing the computation of the one-loop nonzero mode determinants in instanton-monopole backgrounds and the deformation of the moduli space metric are given in Appendices \ref{funddets}, \ref{couplingappx}, \ref{deformation}, and \ref{next}.

 In Section
\ref{antiperiodic}, we study the center-symmetry realization in massive SQCD with all  quark supermultiplets antiperiodic around $\S^1$. Since both quarks and squarks are antiperiodic, this boundary condition preserves supersymmetry. 
In this case,  upon decoupling squarks and gauginos, the theory becomes  thermal QCD with massive quarks. We compute (analytically or numerically, depending on the quark mass) the deviation of the Polyakov loop expectation value from the center-symmetric value. We 
 show that  the semiclassical calculation is self consistent down to values of the Dirac mass (at least) of order the dual photon mass in the given vacuum. We also argue (see also Section \ref{qualitative}) that, for sufficiently massive quarks, there is a crossover of the Polyakov loop from almost center-symmetric to center-broken upon varying $m_{soft}$ and $L$, as in  thermal lattice QCD (simulations are always performed with massive quarks, see \cite{Heller:1984eq} for an early reference).
  
 In Section 
 \ref{centersymmetricquarks}, we study the  boundary conditions for quarks around $\S^1$ which allow SQCD to have an exact center symmetry for any scale of the Dirac mass (but with particular flavor symmetry). Unlike the case of all-antiperiodic quarks, upon decoupling superpartners, this theory does not acquire a thermal interpretation,  but it allows us to study the topological objects for a broader range of values of the  quark masses.  We study in some detail two representative cases.
 We argue that in one of these cases, Section \ref{case1}, the semiclassical calculation is valid down to arbitrarily small finite Dirac mass and that center symmetry is preserved   
 (to substantiate this, we also present an exact superpotential describing the moduli space of the massless theory away from the origin and argue that in the limit of zero center-symmetric Dirac mass the theory is in the center-symmetric vacuum on this moduli space).
 The other case with exact center symmetry, Section \ref{case2}, allows for arbitrarily light physical mass of quarks in the center-symmetric vacuum, but we show that semi-classics breaks down before the massless limit is achieved. 
 
In Section 
  \ref{bionpotentialsection}, we write  the potential for the Polyakov loop and the dual photon for all cases mentioned above. As in pure SYM, the potential is due to magnetic and center-stabilizing bions. We argue that in all cases considered, when semiclassics is applicable, quarks only contribute to the bion dynamics by deforming the monopole-instantons due to their quantum fluctuations. In cases when the physical mass of the quarks in the given vacuum (arising from  both the holonomy and the Dirac mass) can be made smaller than the inverse bion size---i.e., exactly when one would expect quarks to contribute to the instanton-antiinstanton binding---the semiclassical approximation and the associated bion picture of the vacuum can be seen to break down. 
  
  In Section
  \ref{conclusion}, we briefly summarize and mention prospects for future work.

\section{Review of the phase diagram of softly-broken pure  SYM on $\R^3 \times \S^1$}
 \label{symreview}
 
 The benefits of studying the theory on $\R^3\times \S^1$ are several. Due to the abelianization of the gauge group along the Coulomb branch, the coupling is perturbative at small $L$. The three dimensional nature of the long-distance theory also allows a dual description of the magnetic sector in the effective compact $U(1)$ theory  via an abelian dual compact scalar field $\sigma$ first introduced in \cite{Polyakov:1976fu}. On the Coulomb branch of an $SU(2)$ theory, the relevant long-distance fields are  the dual photon $\sigma$ and the component of the gauge field $A_3^3$ along the Cartan generator ($\tau^3\over 2$) and the compactified direction (labeled here by $x_3$). The latter effectively describes the expectation value of the holonomy along $\S^1$. Instead of using the expectation value of $A_3$ to label the Coulomb branch, it is convenient to introduce a dimensionless field $b'$ which denotes its deviation from the center-symmetric value.\footnote{For its precise definition see (\ref{bprime}); $b'=0$ corresponds to the center-symmetric holonomy.}
  
 In addition to perturbative fluctuations, in the case of pure SYM with $N_c=2$ the fundamental topological objects that contribute to the dynamics are the BPS and KK instanton-monopoles.\footnote{These two types of instanton-monopoles are, in our view, most clearly introduced via D-branes, see \cite{Lee:1997vp}, although there are also   field-theoretical descriptions, notably \cite{Kraan:1998pm}. See \cite{Anber:2011de} for a recent introduction and many references.} They generate 't Hooft operators which we, schematically, denote as:
\begin{eqnarray}
\label{susymonopoles}
&&{\cal M}_{1} =   e^{- S_0} e^{ - b' +i \sigma}     \bar\lambda \bar\lambda, \qquad 
  {\cal M}_{2} =e^{- S_0}   e^{ + b' -i \sigma} \bar\lambda \bar\lambda,  \cr  \cr
&&\overline {\cal M}_{1} =  e^{- S_0}  e^{ - b' - i \sigma} 
                             \lambda  \lambda, \qquad 
  \overline {\cal M}_{2} = e^{- S_0}   e^{ + b' + i \sigma}
                              \lambda \lambda, 
\end{eqnarray}We use  ${\cal M}_1$ to denote the BPS monopoles and ${\cal M}_2$---the KK monopoles, and $e^{-S_0} = e^{ -{ 4 \pi^2 \over g^2}}$ is the 't Hooft suppression factor at the center symmetric point $b'=0$. Finally, the fields $\lambda$ are the components of the gauginos along the Cartan direction. 
In the supersymmetric limit, the 
long-distance theory is described by:
\begin{eqnarray}
\label{lagr2}
{\cal L}_{SUSY}=&&  {1 \over 2} {g^2 \over (4 \pi)^2 L}
  \left[(\partial_i b')^2 + (\partial_i \sigma)^2\right] 
 +  i  \frac{L}{g^2} \bar \lambda \bar\sigma^i \partial_i \lambda  
 +    \alpha  e^{- {4 \pi^2 \over g^2}} \left[ 
   \left( e^{-b'+i \sigma} +  e^{+b'-i \sigma}\right ) \bar\lambda \bar\lambda  
 +  {\rm c.c.}  \right] 
 \cr \cr
&& \mbox{}+ \beta \; \frac{e^{-   {8 \pi^2 \over g^2}}}{L^3}  
   \left[ e^{- 2b'} +  e^{ 2b'} -   e^{- 2i \sigma }  -  e^{ 2i \sigma } 
\right]   ~.
\end{eqnarray} The spinor notation here and below is that of \cite{Wess:1992cp} (the index $i = 0,1,2$ and $\bar\sigma^i$ are the first three $\bar\sigma$ matrices of that reference).
 All couplings in (\ref{lagr2}) are normalized at the scale $L^{-1}$ and only the exponential dependence on $g^2$ is kept in the nonperturbative terms; the power law dependence on  $g^2$  is hidden in the coefficients $\alpha, \beta$.
The physics giving rise to (\ref{lagr2}) is as follows: 
\begin{enumerate}
\item The kinetic terms are the dimensional reduction of the 4d kinetic terms (for $b'$ and $\lambda$); while the term containing $\sigma$ is obtained after a photon---dual photon duality transformation.
\item The term proportional to $\alpha$ is due to the 't Hooft vertices of the monopole-instantons ${\cal{M}}_{1,2}$ along with their zero modes, see (\ref{susymonopoles}). 
\item The origin of the terms in the scalar potential deserves some more explanation. As usual in supersymmetry, the  potential term in (\ref{lagr2}) can be derived from a holomorphic superpotential  to be discussed in later Sections. However,  we will now pause and discuss a different way of thinking about the scalar potential terms, as the physics behind these terms transcends supersymmetry:\begin{enumerate}
\item
The terms giving rise to the $\sigma$ potential ($- \cos 2 \sigma$) are due to ``magnetic bions"---correlated instanton-antiinstanton events of the type ${\cal{M}}_1$-$\overline{\cal M}_{2}$ and $ {\cal{M}}_2$-$\overline {\cal M}_{1}$. This type of instanton ``molecules"   have no topological charge, but carry magnetic charge under the unbroken $U(1)$ and are responsible for confinement. The amplitude associated with an $[{\cal M}_{1} \overline {\cal M}_{2} ]$ 
composite is:
\begin{eqnarray} 
 [{\cal M}_{1} \overline {\cal M}_{2} ] \sim 
 {\cal A} e^{-2S_0} e^{2i \sigma},  
\end{eqnarray}
where
\begin{eqnarray} 
\label{lambda1}
 {\cal A} =   \int d^3r\,  
     e^{- \left(  2 \times \frac{4\pi L}{g^2  r} + 4  \log r \right)   }   
=  4 \pi I (\lambda)~,\qquad  \lambda \equiv {g^2 \over 8 \pi L}~
\end{eqnarray}
denotes the integral over the quasi-zero mode (the separation between the instanton-monopole constituents). 
The meaning of the terms in the exponent is as follows: $2 \times 
\frac{4\pi L}{g^2  r}$ accounts for the repulsion due to exchange of 
$\sigma$ and $b$-scalars, and $4  \log r$ is the attraction due to gluino
fermion zero mode exchange. Consequently, there is a single saddle-point 
in the  quasi-zero mode integral, given by:
\begin{equation}
r_{\rm b} = \frac{4 \pi L}{g^2}~,
\label{bionsize}
\end{equation}
which can be interpreted as the magnetic bion size. The bion size
is much larger than monopole-size, but much smaller than (uncorrelated) 
inter-monopole separation. Consequently, a representation of the partition 
function as a dilute gas of magnetic bions is justified. 
\item The terms  giving rise to the $b'$ potential ($\cosh 2 b'$) are due to ``center stabilizing neutral bions"---correlated instanton-antiinstanton events of the type 
${\cal{M}}_1$-$\overline {\cal M}_{1}$ and ${\cal{M}}_2$-$\overline {\cal M}_{2}$. These types of correlated tunneling events have no topological charge, but have scalar ``charge", i.e., couple to the holonomy $b'$.  Here, the integral over the quasi-zero mode is, 
naively:
\begin{eqnarray} 
[{\cal M}_{1} \overline {\cal M}_{1} ] \sim 
 {\cal A}_{\rm naive} e^{-2S_0} e^{- 2b' } .
\end{eqnarray}
Now, the interactions  between constituents due to $\sigma$ and $b$ 
exchange are both attractive, while the fermion zero mode induced 
attraction is not altered (it remains attractive). Thus the correlated amplitude is:
\begin{align}
 {\cal A}_{\rm naive}(g^2) = \int d^3r   \;        
 e^{- \left(  - 2 \times \frac{4\pi}{g^2  r}  + 4  \log r \right)   }   
&=  4 \pi  \tilde I(\lambda) \nonumber \\ 
& \rightarrow \;\; {\cal A}_{\rm BZJ}(g^2) =  - 4 \pi I(\lambda).
\label{naive}
\end{align}
 The integral on the first line in  
(\ref{naive}), as it stands, is dominated by the small $r$ regime, 
where not only (\ref{naive}) is incorrect, it is also hard to make 
sense of constituents as the interaction becomes large (this is, clearly, in 
sharp contrast with the magnetic bion
\cite{Unsal:2007jx,Anber:2011de}).   However, this ``molecular" interpretation can be made sensible by analytic continuation in $\lambda$, as indicated by the arrow  in the second line in (\ref{naive}) (see Refs.~\cite{Poppitz:2011wy,Argyres:2012vv,Poppitz:2012sw} for more details on $ {\cal A}_{\rm BZJ}$). We only note that the phase (sign) difference between ${\cal A}_{\rm BZJ}$ in Eq.~(\ref{naive}) and ${\cal A}$ in Eq.~(\ref{lambda1}), obtained by analytic continuation, coincides with the one following from supersymmetry. 
\end{enumerate}
\end{enumerate}
 When a small soft mass $m_{soft}$ for the gauginos on $\R^3 \times \S^1$ is introduced,  extra terms appear in the Lagrangian (\ref{lagr2}), which thus becomes:
\begin{eqnarray}
\label{lagr3}
{\cal L}_{SUSY + soft}&& =\nonumber \\
&&   {g^2 \over 2 (4 \pi)^2 L}
  \left[(\partial_i b')^2 + (\partial_i \sigma)^2\right] 
 +  i  \frac{L}{g^2} \bar \lambda \bar\sigma^i \partial_i \lambda  
 +    \alpha  e^{- {4 \pi^2 \over g^2}} \left[ 
   \left( e^{-b'+i \sigma} +  e^{+b'-i \sigma}\right ) \bar\lambda \bar\lambda  
 +  {\rm c.c.}  \right] 
 \cr \cr
&& \mbox{}+ 2 \beta \; \frac{e^{-   {8 \pi^2 \over g^2}}}{L^3}  
   \left[\cosh 2 b' -   \cos 2 \sigma \right]   \\
&&  +\: {m_{soft} \over g^2} \;  (\lambda \lambda + {\rm c.c.}  )  -  c_1 {m_{soft}^2\over L}  b'^2 +  \gamma {m_{soft} \over L^2}  e^{- {4 \pi^2 \over g^2}} 
   \left( e^{-b'+i \sigma} +  e^{+b'-i \sigma} + {\rm c.c.} \right) ~.\nonumber
  \end{eqnarray}
The origin of the new terms (on the last line above) is easy to explain. The first term is simply the gaugino mass term. The second term is due to the one-loop perturbative potential for the holonomy  \cite{Gross:1980br}, which introduces a center-destabilizing potential for $b'$ (the term given in (\ref{lagr2}) is the leading term in a small $m_{soft}$ expansion of the holonomy  potential of \cite{Gross:1980br}).  The third term is due to the lifting of monopole-instanton zero modes by the gaugino mass (its form and the coefficient $\gamma$ can be obtained by contracting the zero modes in the $\alpha$-term in (\ref{lagr2})). 

As shown in \cite{Poppitz:2012sw}, the terms with coefficients $\beta$ and $\gamma$ in (\ref{lagr3}) are the leading ones that determine the phase structure of the pure softly-broken SYM theory in the small-$L$, small-$m_{soft}$ domain. There is a center-symmetry-breaking quantum phase transition as the dimensionless parameter $m_{soft}Le^{S_0}$ is varied (by writing the potential in (\ref{lagr3}) in dimensionless terms, it is clear that $m_{soft}Le^{S_0}$ controls the relative strength of the supersymmetry-breaking terms in the potential). Notice that  $m_{soft}Le^{S_0}$ can equivalently be written as $m_{soft}\over L^2 \Lambda^3$ and that we keep the strong coupling scale $\Lambda$ fixed (as ($\Lambda L)^3 = e^{- S_0}$). Thus, temporarily setting $\Lambda = 1$, at small values of $m_{soft}\over L^2$, the center-stabilizing neutral bions dominate and the ground state is center symmetric. At $m_{soft} \over L^2$ larger than some critical value, the $\gamma$-term in (\ref{lagr3}), due to monopole-instantons, destabilizes the center-symmetric vacuum and leads to a second-order (for $N_c=2$) center-symmetry breaking transition. This is also the known order of the deconfinement transition in nonsupersymmetric thermal $SU(2)$ YM theory.

 Further evidence for the continuous connection of the small-$m_{soft}$, small-$L$ center-breaking transition to the thermal deconfinement transition in YM theory was given in \cite{Poppitz:2012nz}. 
For $N_c>2$ a first-order transition was found, as  seen on the lattice in thermal pure YM theory, see the recent review \cite{Lucini:2012gg}. The phase transition temperature also acquires  topological $\theta$-angle dependence  due to the ``topological interference" discussed in \cite{Unsal:2012zj}; see also \cite{Parnachev:2008fy, Thomas:2011ee} for earlier related discussions of $\theta$-dependence. The $\theta$-dependence of the critical $L_c$ was  studied in \cite{Poppitz:2012nz,Anber:2013sga} and is  in  qualitative agreement with  recent lattice studies in thermal pure YM theory, see 
\cite{D'Elia:2012vv, D'Elia:2013eua} and references therein.\footnote{ We note that Ref.~\cite{Anber:2013sga} also studied the $\theta$-dependence of the discontinuity of the Polyakov loop trace at the transition in softly broken SYM and found a dependence confirmed by  the recent lattice studies of Ref.~\cite{D'Elia:2013eua}.}
We also note that a model of the deconfinement transition in pure YM, incorporating the center-stabilizing neutral-bions in an instanton-monopole liquid, along with a comparison with lattice data, was recently
proposed in \cite{Shuryak:2013tka}.

Finally, we mention the case most closely related to the present paper. This is the study  \cite{Poppitz:2012nz} of a   theory without center symmetry: pure SYM with gauge group $G_2$ and soft-breaking on $\R^3 \times \S^1$. This theory is similar to real QCD in that fundamental  quarks  can be screened (in $G_2$, by three gluons). Proceeding along the lines described above for $SU(2)$,  one finds  a discontinuous transition of the Polyakov loop from almost center-symmetric to center-breaking upon increasing  $1\over L$, as seen on the lattice for  thermal   $G_2$ YM theory
\cite{Pepe:2006er,Cossu:2007dk}.

\section{Towards QCD: adding fundamental flavors}
\label{towards}

  We now generalize the setup of Ref.~\cite{Poppitz:2012sw} by adding $N_f$ massive chiral superfields in the fundamental of the gauge group (their fermionic parts constitute $N_f$ Dirac fundamental flavors). This introduces $N_f$ fundamental quark flavors and $2N_f$ complex scalars (thus, each quark flavor comes with two complex fundamental scalars, in the fundamental and 
  antifundamental representation, respectively), with the action: 
\begin{equation}
\label{lagrfund}
S_{fund}=\int_{\R^3\times \S^1}\; \left(\bar\psi (i\slashed D-iM)\psi+ |D_\mu\phi|^2+|\phi|^2M^2\right)+\ldots
\end{equation}
where $\slashed D=\gamma^\mu D_\mu, \mu=0,1,2,3$, and $\gamma^\mu$ are four dimensional  gamma matrices, where we have momentarily departed from the notation of \cite{Wess:1992cp} (the dots denote additional interaction terms in the SQCD  action, the superpartners of the interaction with the gauge boson). We assume that each quark flavor  multiplet has an arbitrary periodicity in the compact direction
\begin{equation}
\label{boundary}
\psi(\vec x,x_3+L)=e^{i\alpha}\psi(\vec x,x_3)\;,\qquad\phi(\vec x,x_3+L)=e^{i\alpha}\phi(\vec x,x_3)\;,  \qquad\alpha  \equiv  uL~.\end{equation}
Equivalently, $\alpha$ shows up as a constant $x_3$-component $u$ of a $U(1)$ gauge field $A_3\rightarrow A_3+u$, with $u=\alpha/L$. In the supersymmetry literature this is sometimes referred to as ``real mass'' (notice that the background field $u$  gauges a $U(1)$ subgroup  of the vectorlike global flavor symmetry). Since both components of the matter superfields have the same boundary condition, this boundary condition preserves supersymmetry.  

In our further study, we consider several different choices\footnote{In this paper, we will not study  in detail the Polyakov loop crossover for   generic values of $uL$. We only note that these correspond to introducing imaginary chemical potential for baryon number and that corresponding lattice studies exist  \cite{Chen:2004tb}.}
 of $\alpha$. For the purpose of interpolating to thermal QCD, we note that taking $\alpha=\pi$ for all matter multiplets makes them antiperiodic around the compact direction. Thus, when soft supersymmetry breaking terms are added (gaugino and scalar masses of order $m_{soft}$), one expects that upon making the scalars $\phi$ and gauginos massive, this theory  interpolates between the supersymmetric theory with a fundamental supermultiplet on $\R^3 \times \S^1$, at $m_{soft}=0$, and  thermal QCD with fundamental quarks, at $m_{soft}\rightarrow \infty$.

For completeness, we note that the vector multiplet part of the action is that of pure SYM (again, omitting the Yukawa interactions of the gauginos with the squarks and quarks):
\begin{equation}
S_{vec}=\int_{\R^3\times \S^1}\frac{2}{g_4^2}\Tr\left(\frac{1}{4}F_{\mu\nu}^2+i\bar\lambda\bar \sigma^\mu D_\mu\lambda+c.c.\right)
\end{equation}
where $\lambda=\lambda^a t^a, F_{\mu\nu}=F_{\mu\nu}t^a$ with $t^a=\frac{\tau^a}{2}$ and $\tau^a$ are the usual Pauli matrices. 
In pure SYM, upon compactification, due to the SU(2)$\rightarrow$ U(1) breaking by the compact Higgs field $A_3$, the tree-level low-energy effective action of the vector multiplet was given by:
\begin{equation}
\label{lagr33}
S_{vec}^{eff}=\int_{\R^3} \left(\frac{L}{4 g_4^2}{F_{ij}}^2 +{L\over2 g_4^2} (\partial_i v)^2 + i\bar\lambda^3\bar \sigma^i \partial_i\lambda^3+c.c.\right)~,
\end{equation}
where $F_{ij} = \partial_{[ i} A_{j ]}^3$ and we have selected a gauge such that the Higgs field is:
\begin{equation}
\label{higgs}
A_3=v\; \frac{\tau^3}{2}~.
\end{equation} The components of the gauge field and gaugino that become massive, with mass of order $v$, are omitted from (\ref{lagr33}). Notice  that  the kinetic terms given in 
(\ref{lagr2}) are exactly the terms in (\ref{lagr33}) upon a duality transformation $\partial_i \sigma \sim \epsilon_{ijk}F^{jk}$, relabeling $\lambda^3 \rightarrow \lambda$, and redefining $v \rightarrow b'$ as in (\ref{bprime}).

For $v\ne 0$, i.e., along the Coulomb branch, 
the fundamental matter is always massive and should be integrated out, except in the region where the real mass $u\sim  {v_{min} \over 2}$, where $v_{min}$ is the vev of the Higgs field (\ref{higgs}), and when $M$ is small, see Sec.~\ref{case2}. The perturbative effect of the fundamental multiplet on the holonomy is, as  usual, canceling between the superpartners. However in the background of the instanton-monopoles this cancelation is not complete (but it is exactly calculable, see Appendix \ref{funddets}). Upon the introduction of SUSY breaking mass to the scalars, however, even the perturbative contribution influences the holonomy, and, as is well known, generically prefers trivial holonomy. Its effect is similar to the one of the gaugino mass in pure SYM, the term $\sim c_1$ in (\ref{lagr3}), and will be described in Section \ref{qualitative} (see Appendix \ref{taming} for explicit formulae). 
 
 \subsection{Symmetries and monopole-instanton zero modes}
 
\label{fundsymm}

Our goal in this paper is to study the fate of the center-symmetry breaking transition in SYM at small $m_{soft}$ and $L$ (described in Section \ref{symreview}) when fundamental fermions are added. We expect that the transition is turned into a crossover, as observed in lattice studies of QCD with dynamical fundamental fermions. We would like to see how this is manifested in the calculable picture on $\R^3 \times \S^1$. 
It is thus mandatory that we consider the symmetries of the theory with fundamental matter.

Note that in the $SU(2)$ theory that we consider here, the flavor symmetry of   massless SQCD is  $U(1)_A \times SU(2 N_f)$, as the matter fermions can be taken to be $2 N_f$ fundamental Weyl fermions (unlike in Eq.~(\ref{lagrfund}); in this notation, the $U(1)$-baryon global symmetry mentioned above is embedded in $SU(2N_f)$ and is generated by the Cartan generator  $\sim {\rm diag} (1,  \ldots 1, -1,  \ldots -1)$). The Dirac mass term is given by a  $2 N_f \times 2 N_f$ antisymmetric matrix $M^{ij}$ and has the form:
\begin{equation}
\label{diracmass}
L_{Dirac} = M^{ij} \psi_i^{\alpha a} \psi_{\alpha \; j}^b \epsilon_{ab} + {\rm h.c.},
\end{equation}
where $\alpha$ are the $SL(2,C)$ spinor indices, $a,b$---the fundamental $SU(2)$ indices, and $i,j$---the $2N_f$ flavor indices. Furthermore, notice that, in addition to the Dirac mass, which explicitly breaks the chiral symmetry (e.g.,   when $M^{ij} \sim J^{ij}$: $SU(2N_f)\rightarrow SP(2 N_f)$), the fermions also have a real mass due to the expectation value of $A_3$ along the Coulomb branch (for example, in the center-symmetric vacuum, an $SU(2)$ doublet splits into two fermions of opposite  $U(1)$ charges and mass $ \pm {\pi\over 2L}$, respectively).

It is known, from the index theorem for the Dirac operator on $\R^3 \times \S^1$, see \cite{ Poppitz:2008hr} and references therein, that fundamental zero modes localize to one of the two types of monopole instantons (in the center symmetric vacuum with $v  ={\pi\over L}$, and for periodic fundamental fermions, this would be the ${\cal{M}}_1$ monopole instanton; 
for antiperiodic fermions, this would be the ${\cal{M}}_2$). Thus, instead of (\ref{susymonopoles}), we expect to have the ``monopole operators":\begin{eqnarray}
\label{susymonopoles1}
&&{\cal M}_{1} =  e^{-S_0} e^{ - b' +i \sigma}     \bar\lambda \bar\lambda \bar\psi^{2 N_f}, \qquad 
  {\cal M}_{2} = e^{-S_0}   e^{ + b' -i \sigma} \bar\lambda \bar\lambda,  \cr  \cr
&&\overline {\cal M}_{1} = e^{-S_0} e^{ - b' - i \sigma} 
                              \lambda \lambda \psi^{2 N_f}, \qquad 
  \overline {\cal M}_{2} = e^{-S_0}   e^{ + b' + i \sigma}
                              \lambda \lambda, 
\end{eqnarray}
where $\psi^{2 N_f} = {\rm Pf} (\psi \cdot \psi) $ denotes the flavor-symmetric 't Hooft determinant.\footnote{Recall  the Pfaffian of an antisymmetric matrix $X$: ${\rm Pf} X$$=$$\epsilon^{i_1 i_2 i_3 i_4 \ldots i_{2 N_f -1} i_{2 N_f}} X_{i_1 i_2} X_{i_3 i_4} \ldots X_{i_{2 N_f - 1} i_{2 N_f}}$.}
The meaning of the ``monopole operators"   in (\ref{susymonopoles1})  needs to be  precisely defined, in view of the facts that: 
{\it i.)}
The mass of $\psi$ in the center-symmetric vacuum is of order $ {1\over L}$ and it is not clear why $\psi$ should appear in the effective theory at scales larger than $L$. {\it ii.)}
 Even if a component of the $\psi$ doublet is kept lighter than ${1\over L}$ by turning on a nonzero $u$, recall (\ref{boundary}), it still carries electric $U(1)$ charge. Hence, $\psi$ is a ``magnetic" object from the point of view of the dual photon $\sigma$, needed to write a local monopole operator. Thus, writing a lagrangian containing terms like ${\cal M}_1$, with both electric and magnetic fields, requires  more elaboration. Furthermore, one expects that a light electrically charged $\psi$ would drive the $U(1)$ coupling strong and invalidate the semiclassical analysis.

 Leaving aside these important subtleties (as we shall see, the ``power of supersymmetry" will, without mercy, force us to return to them), for now we shall adopt the  view that Eq.~(\ref{susymonopoles1})  provides a useful book-keeping device for encoding symmetries. We shall now attempt, via symmetry and simple dynamical arguments, to guess a form for the resulting potential for $b'$ and $\sigma$ in the theory with fundamentals---assuming an Abelianized description is valid. 

We begin by discussing the chiral symmetries and their reflection in the monopole operators (\ref{susymonopoles1}). The nonabelian  $SU(2N_f)$ symmetry is intact (at small $\S^1$, it is only explicitly broken by the mass term (\ref{diracmass})). On the other hand,  the two classical $U(1)$ chiral symmetries (see (\ref{3dsym}))---$U(1)_\lambda$, acting by a phase on $\lambda$, and $U(1)_A$, acting by a phase on all $\psi_i$---are anomalous. Only a linear combination $U(1)_X$: $\lambda \rightarrow e^{- i \omega} \lambda$, $\psi_i \rightarrow e^{ i {2 \omega \over  N_f} } \psi_i$ is anomaly free. Clearly the 4d BPST-instanton 't Hooft vertex $\sim \lambda^4 \psi^{2 N_f}$ is invariant. On the other hand, the monopole operators (\ref{susymonopoles1}) are invariant only up to a shift of the dual photon field, i.e., $U(1)_X$: $\sigma \rightarrow \sigma + 2 \omega$. The intertwining of the anomaly free chiral symmetries and the topological shift symmetry of the dual photon is a generic feature in theories on $\R^3 \times \S^1$ \cite{Aharony:1997bx} and is summarized  below:
 \begin{equation}
\begin{array}{cccc}
&  U(1)_{\lambda}   & U(1)_A & U(1)_X \cr
\lambda & 1 & 0 & -1 \cr
\psi_i & 0 & 1 & {2 \over N_f}\cr
e^{i \sigma} &  -   & -  & 2\cr 
\end{array}~~.
\label{3dsym}
 \end{equation} 

The anomaly-free $U(1)_X$, acting on both the gaugino and fundamental fermions, is explicitly broken when either a Dirac mass term $M$ or a  gaugino mass term $m_{soft}$ are introduced. However, in the absence of both $M$ and $m_{soft}$,  $U(1)_X$ is an exact symmetry and a potential for the dual photon is not allowed (hence, the $M=0$ supersymmetric theory would  not be expected to confine, at least in the calculable semiclassical regime, assuming an abelianized dual photon description is appropriate---for an example of such a non-confining theory on the Coulomb branch on $\R^3 \times \S^1$, see Section \ref{case1}).
On the other hand, chiral symmetry alone does not forbid a potential for $b'$ (although when $m_{soft}=0$, the constraints of supersymmetry have to be obeyed---in the supersymmetric limit any potential should arise from a superpotential, a requirement which can be quite restrictive and will be considered further below).  
 Thus, the monopole instantons (\ref{susymonopoles1}) might be expected to contribute to the scalar potential for $b'$ via terms similar to those in (\ref{lagr2}) even when $M=0$. The statements in this paragraph were all made without taking any constraints of supersymmetry into account; we shall see that these are strong and can modify the naive picture.

Finally, note that the symmetry between the contributions of ${\cal M}_1$ and ${\cal M}_2$ in pure SYM theory is broken when fundamental fermions are introduced, as they ``prefer" to localize zero modes at one of the fundamental monopole instantons. It is indeed the ${\cal M}_1$ $\leftrightarrow$ ${\cal M}_2$ symmetry that is responsible for the nonperturbative stability of the center-symmetric Coulomb branch vacuum ($b'=0$) in the SYM case. As center symmetry is  usually absent with fundamental matter, the ${\cal M}_1$ $\leftrightarrow$ ${\cal M}_2$ symmetry is also not present, hence (generically---we shall see that there are exceptions) an exactly center-symmetric vacuum is not expected at any $L$.
  
\subsection{Qualitative expectation for (approximate) center symmetry in softly-broken massive SQCD and  Polyakov loop crossover }
\label{qualitative}

Let us continue, still without due worry about the constraints of supersymmetry, and consider what one expects along the Coulomb branch of the theory, but now with  massive quark supermultiplets---so that we might lift the zero modes from Eq.~(\ref{susymonopoles1}) and not worry about the subtleties mentioned thereafter---with regards to the role of the various fundamental topological objects. 
 The neutral bions of ${\cal M}_1$-$\overline{\cal M}_1$ type are expected to be present, but should be affected by the presence of fundamental fermions (we assume zero modes localize on ${\cal M}_1$). 
 As a first approximation, it appears that the neutral bions of ${\cal M}_2$-$\overline{\cal M}_2$ should not be significantly affected.
   Magnetic bions  ${\cal M}_1$-$\overline{\cal M}_2$ are also expected to  form in the massive Dirac theory and should be affected by the quarks as well. 
Finally,  introducing soft masses $m_{soft}$ would lift gaugino zero modes and give a  ${\cal M}_{1,2}$ contribution to the potential, as in (\ref{lagr3}). Of course, one expects that ${\cal M}_1$ and  ${\cal M}_2$ are affected differently by the fundamentals as in the limit of vanishing $M$, one carries a fundamental zero mode and the other does not.

In this Section, we will take the liberty to mix qualitative expectations,  determining the form of the holonomy potential, with some results of the calculations of later Sections, especially the scaling of the parameters of the potential.
Continuing with the qualitative discussion,   
provided the abelianized description is valid for some range of  $M\ne0$ (and soft masses $m_{soft} \ne 0$), we propose a scalar potential, quite similar to the one that appears in the pure SYM case of Eq.~(\ref{lagr3}): 
   \begin{align}
 \label{zerodiracpotential1}
{ e^{2 S_0} \over (ML)^{N_f} }L^3 V(\sigma, b') &\sim  \cosh( 2  b' - \delta_1)   + c_1^\prime \cos 2 \sigma \nonumber \\
 &
 +  c_3^\prime {m_{soft} L e^{S_0} \over (LM)^{N_f\over 2}} \; \cosh(b' - \delta_2) \cos\sigma - c_5^\prime \;  {m_{soft}^2 L^2 e^{2 S_0} \over (LM)^{N_f} }\;  b'^2 ~.
 \end{align} 
The coefficients $\delta_1$ and $\delta_2$ express our expectation that different bions will be affected differently by the presence of fundamental fermions. 
 The overall normalization in the softly-broken SQCD scalar potential (\ref{zerodiracpotential1}) follows from the calculation in Section \ref{bionpotentialsection}.  The soft terms on the last line above are also similar to the ones in (\ref{lagr3}) and represent the leading effect of soft breaking. The dimensionless coefficient  appearing in the  soft-breaking  terms  can be expressed via renormalization group invariant quantities as: 
\begin{equation}
\label{cexpression}
c\equiv c_3'm_{soft}Le^{S_0} (LM)^{-{N_f\over 2}}\sim c_3' m_{soft}L (L\Lambda)^{-3+{N_f \over 2}} (ML)^{-{ N_f \over2}} = c_3' \; {m_{soft} \over L^2} \; {\Lambda^{-3 + {N_f \over 2}}   M^{-{N_f \over 2}}}~,
\end{equation}
where $\Lambda$ is the one-loop strong coupling scale of the $N_f$-flavor theory (note that $\Lambda^{-3 + {N_f \over 2}} M^{-{N_f \over 2}}$ $\rightarrow$ $\Lambda_{SYM}^3$ as $M$ becomes large). 
At this stage we do not yet know the precise values of the coefficients $c_i'$ and $\delta_i$, but we expect them to depend on $ML$ and $N_f$, and to approach the values of the pure SYM theory when $M\rightarrow \infty$; see Section \ref{susyvacuum}.
 
It is clear that for small soft breaking (i.e., small $c$ of Eq.~(\ref{cexpression})), the minimum for $b'$ is determined by the  terms on the first line of (\ref{zerodiracpotential1}) and is shifted away from $b'=0$ (the center symmetric value) to $b' =  \delta_1/2$ (thus, to a center-symmetry breaking minimum), while $\sigma$ is minimized at $0$ or $\pi$ (for positive $c_1^\prime$; note that  the $c_3'$ term will select one of these values). Naturally, the absence of center symmetry at any ``temperature" ($1/L$)  is  expected in the theory with fundamentals.  However, for sufficiently heavy quarks,   the deviation from center symmetry at small-$L$ (for fixed $m_{soft}$) should be small and this should be reflected in  $\delta_{1,2}$. Increasing the soft mass ($c$) destabilizes the small-$b'$ minimum and shifts it away towards a value when the abelian picture breaks down, as in the softly-broken pure SYM case. We interpret this as turning the smooth (2nd order) center-symmetry breaking transition in SYM with increasing the $c$
parameter into a crossover---as observed in lattice simulations of QCD with massive quarks.

In the remaining Sections, we shall, by an explicit calculation verify the qualitative picture described in (\ref{zerodiracpotential1}) in the case of sufficiently massive quarks---such that in the supersymmetric limit, the deviation from center symmetry is parametrically suppressed. For this case, in Section \ref{bionpotentialsection} we obtain analytic expressions for the various coefficients in (\ref{zerodiracpotential1}). Most importantly,  we will  show  that $\delta_1 = 2 \delta_2$ and $c_1^\prime = -1$ (i.e., the dual photon potential has its supersymmetric value, up to small corrections suppressed by powers of $m_{soft}L$, which we neglect).\footnote{See Eq.~(\ref{allantipotential}) for the potential and  Eq.~(\ref{fermionbilinear1}) for the fermion bilinear term in the potential (which,  upon soft-breaking, gives rise to the monopole-instanton term in (\ref{zerodiracpotential1})). } The coefficient of the monopole-instanton term, $c_3^\prime$, can be computed as in \cite{Poppitz:2012sw} and its pre-exponential $g^2$-dependence (which we do not show here
for brevity\footnote{Here, we also do not give  the explicit expressions for $\delta_{1}=2\delta_2$, $c_3'$, and $c_5'$,  as they are not important for the qualitative discussion of the transition ($c_3'$ and $c_5'$ have pre-exponential $g^2$ dependence which we omit, see \cite{Poppitz:2012sw}); for an explicit expression for $\delta_2$ in the heavy-quark limit, see Eq.~(\ref{bprimenoncenter}).})  implies that it is dominant w.r.t.~the perturbative one-loop contribution. Thus,   in the weakly-coupled semiclassical regime, {\it the almost center-symmetric/center-breaking  crossover  is largely due to a competition between neutral bions and monopole-instantons.} 

Now it is easy to see that the minimum of the potential for $b'$ is at $b' =  {\delta_2}$ for sufficiently small values of $m_{soft}$, while for $c>4$ it is given by $b'_{min}=\delta_2\pm \cosh^{-1}\frac{c}{4}\approx -\delta_2+\sqrt\frac{c-4}{2}$,  which is clearly still second order transition (i.e., the $b_{min}'(c)$ has an infinite first derivative). Note however that the there is no center symmetry (i.e., the potential is not invariant with respect to the change $b'\rightarrow -b'$), but to leading order in the soft SUSY breaking mass there is still a $b'\rightarrow -b'+2\delta_2$ symmetry. The symmetry is broken though by the perturbative term proportional to the gluino mass\footnote{The potential \eqref{zerodiracpotential1} describes the effect of softly breaking supersymmetry by adding a mass to the gauginos $m_{soft}$; however this immediately implies a SUSY breaking mass in the fundamental matter multiplet through one-loop effects, giving a higher mass to the squarks $m_0^2\sim g^2m_{soft}^2$. This would imply the existence of a center breaking term  linear in $b'$ in the potential (which follows from the formulae for the matter contribution to the holonomy potential given in Appendix \ref{taming}), however the coefficient  is $g^2$-suppressed compared to the already subleading contribution of the gaugino mass to the perturbative holonomy potential. Further, the linear term does not change the qualitative picture we outline here. We note in passing that we could tune the squark mass to be such that the perturbative potential of the vector multiplet and the matter multiplet combine into the same form as the instanton-monopole terms $(b'-\delta_2)^2$, which would restore the $b'\rightarrow -b'+2\delta_2$ symmetry and again induce a second order phase transition. Although this transition seems of limited interest, it could be relevant for the exploration of the imaginary chemical potential of the phase diagram and its connection to the real chemical potential.} $\sim m_{soft}^2{b'}^2$. Expanding to 4th order in a small quantity $b'-\delta_2$, we find the potential (\ref{zerodiracpotential1}):
\begin{equation}\label{potexpansion}
V(b',\sigma=\pi)\approx \left( {2 \over 3} -\frac{c}{24} \right) (b'-\delta_2 )^4+\left(\frac{4-c}{2}\right)
   (b'-\delta_2 )^2 -c^2 c_5'\; {b'}^2~.
\end{equation}

\begin{figure}[htbp] 
   \centering
   \includegraphics[width=2.5in]{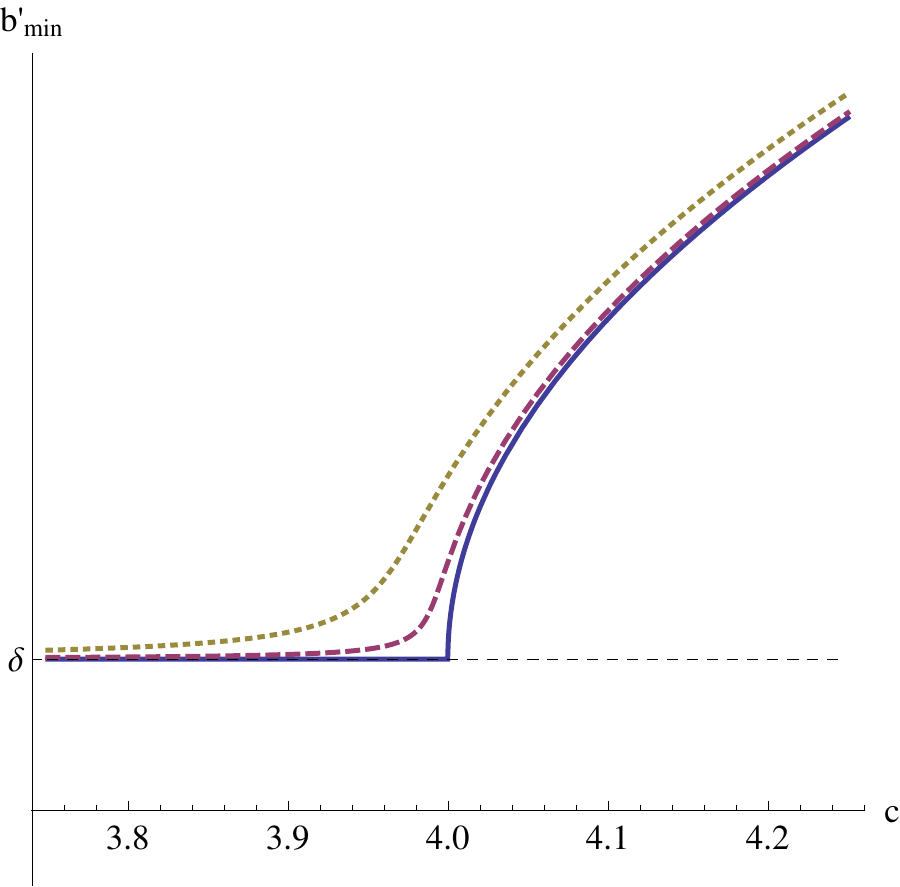} 
   \caption{The minimum of $b'$, proportional to the Polyakov loop trace, $\tr \Omega \approx {g^2 \over 4 \pi} b'$, as a function of the parameter $c= c_3' {m_{soft} \over L^2}\;  \Lambda^{-3+{N_f \over 2}} M^{-{ N_f \over2}}$. 
   The value $b'=\delta$ corresponds to an almost center-symmetric Polyakov loop and the  behavior at $c>4$ indicates a transition towards collapsing eigenvalues of the Polyakov loop.
   The solid-blue, dashed-purple and dotted-yellow curves are for three values of parameter $c_5'=0,0.0001,0.0005$ respectively. Recall that the $c_5'=0$ case corresponds to neglecting the (suppressed by powers of $g^2$) perturbative holonomy potential, leading to a second order transition, which turns into a crossover once $c_5'\ne0$. This is the behavior seen on the lattice, see Ref.~\cite{Heller:1984eq} for an early study of $SU(2)$ theory with dynamical massive quarks.}
   \label{fig:crossover}
\end{figure}

It is easy to see  that taking into account the perturbative potential, the behavior of the minimum  value $b_{min}'(c)$ smooths out and the transition becomes a crossover\footnote{In actuality, there would be another term due to the moduli space metric which would contribute to the monopole term $\sim b'\sinh(b'-\delta_2)$ (see Eq.~(2.34) of reference \cite{Poppitz:2012sw}). This term also does not obey the $b'\rightarrow -b'+2\delta_2$ symmetry, but is subdominant due to the $\sim g^2$ supression. It is however parametrically larger then the ${b'}^2$ term. Nevertheless since the effect of it to leading order is identical to the perturbative ${b'}^2$ we omitted inserting this term in our qualitative expectation of the potential \eqref{zerodiracpotential1}.}, as illustrated in Fig.~\ref{fig:crossover}. Note that since there is a lack of center symmetry (or variation thereof), the Polyakov loop is no longer an order parameter. Nevertheless it is still connected to the energy of an infinitely massive quark, and it shows rapid variation as a function of temperature in lattice QCD simulations.\footnote{It is clear from (\ref{cexpression}) that the fixed-$m_{soft}$ transition from the almost center-symmetric to center-broken phase occurs as $1\over L$ is decreased.}

Finally, recall that the vanishing soft mass limit, $m_{soft}=0$,  is supersymmetric  SQCD with Dirac mass $M$ on $\R^3 \times \S^1$. The dynamics of this theory is not always under analytical (weak coupling) control even at small $L$. Its behavior is qualitatively  known  in the $M=0$ limit and some aspects will be reviewed in the following Section \ref{massive}. On the other hand, the dynamics of SQCD can be quantitatively understood in the limit of sufficiently large $M$, as we will show. It is clear that for large Dirac mass $M$ the theory should go over to pure SYM. For sufficiently large $M$ and any $N_f$, we can 
  understand, in a controlled semiclassical approximation,  the finite-$M$ corrections to the center-breaking transition in pure SYM, and thus its turning to a crossover, due to fundamental flavors.  This will allow us to compute the potential that we guessed in  Eq.~(\ref{zerodiracpotential1}).

 \section{Massive SQCD with $\mathbf{N_f}$ flavors on $\mathbf{\R^3 \times \S^1}$}
\label{massive}

We start by reviewing some exact results.  We will be brief, as there exist both older and more recent extensive literature, see \cite{Aharony:1997bx, Aharony:2013dha} and references therein. 
We only  
consider  $SU(2)$ SQCD with $2 N_f$ doublets in the supersymmetric limit. 
We begin with the massless theory.
 The ``power of holomorphy" in supersymmetry allows one to make exact statements about the superpotential of the theory. Recall that  in the supersymmetric limit every potential has to come from a superpotential, which is strongly constrained, even nonperturbatively, by holomorphy and the symmetries. The exact superpotential of the theory with zero mass matter supermultiplets is given by\footnote{For  $N_f \ge 2$ and all quarks periodic around the $\S^1$ (see \cite{Aharony:1997bx} for other boundary conditions and $N_f=1$). We do not show all cases, since with sufficiently massive quarks, Eq.~(\ref{superpotential0})  describes the low-energy physics in all cases (notice that this also follows from our explicit calculation of Section \ref{susyvacuum}). See Section \ref{case1} for an example of a superpotential with non-periodic quarks and exact center symmetry.}
 \begin{equation} \label{wexact}
 W = (Y {\rm Pf} {\cal{M}})^{1\over N_f - 1} + \eta Y~,
 \end{equation}
 where $Y$ is the ``monopole superfield", which, semiclassically, i.e. sufficiently far form the origin of the Coulomb branch, is $Y \sim e^{ - b' + i \sigma}$, i.e. incorporates the fields $b'$ and $\sigma$ that are classically massless (and superpartners). In (\ref{wexact}), ${\cal{M}}$ does not denote  a monopole operator as in (\ref{susymonopoles}), but the $2N_f \times 2N_f$ meson superfield antisymmetric matrix. The bosonic component of the meson superfield is classically given by ${\cal{M}}_{ij} = \phi_i^a \epsilon_{ab} \phi_j^b$, where $\phi_i^a$ are the squark doublets ($i=1,...2N_f$, $a,b=1,2$). Here, $\eta$ is proportional to the strong-coupling scale of the theory,  $\eta = e^{-{8 \pi^2 \over g^2} + i \theta} \sim \Lambda^{6 - N_f}$ (here $\theta$ denotes the topological theta angle).
 The superpotential (\ref{wexact}) describes the moduli space of the theory---its manifold of ground states, determined by finding the critical point of the $W$ with respect to $Y$ and the components of the meson fields.\footnote{The normalization of the two terms in $W$ is not determined  by symmetry (the normalization is meaningless, unless the K\" ahler potential is also specified)  but this does not diminish its usefulness for studying the moduli space of vacua.}  Classically, the moduli space  consists of a Coulomb branch (where the holonomy has an expectation value)  and a Higgs branch (where the squarks, or the gauge invariant   ${\cal{M}}$) have an expectation value, which meet at the origin (where all vevs vanish). 
 
Quantum mechanically, however, the solutions of $d W=0$ may give rise to a different manifold. In particular, at the singularities of this manifold---such as the one immediately seen from (\ref{wexact}) to occur at $Y={\cal{M}}=0$ for $N_f > 2$---new massless degrees of freedom must occur, leading to a description using a tangled web of dualities. We refer the reader to the  work cited in the beginning of this Section and only briefly discuss the $N_f=2$ case here. Now (\ref{wexact}) has no singularities and hence no   light degrees of freedom in addition to $Y$ and ${\cal{M}}_{ij}$ arise. The vacuum is determined by  $dW=0$: $Y=0$ and ${\rm Pf} {\cal {M}}= - \eta$, i.e., there is a quantum moduli space where the $SU(2N_f = 4)$ chiral symmetry is broken by the expectation value of the lowest component of the meson superfield. Thus, the Coulomb branch vacuum (where the notion of BPS and KK monopole-instantons makes sense) is lifted and the massless theory vacuum is on the Higgs branch. Notice also that, classically,  ${\cal{M}}_{ij}$ is proportional to  $\phi_i^a \epsilon_{ab} \phi_j^b$, whose Pfaffian  vanishes by the rank condition ($a,b=1,2$ and commutativity of $\phi$'s). Thus, the classical Higgs branch ${\rm Pf} {\cal {M}}= 0$ is quantum-deformed to  ${\rm Pf} {\cal {M}}= - \eta$.  One can understand this deformation by studying flows from $N_f=3$ (where a computation using the Seiberg dual is useful) to $N_f=2$, see \cite{Beasley:2004ys}.

The point of the brief review above was  to remind the reader that the physics of massless SQCD on $\R^3 \times \S^1$, even at small $L$, is rather nontrivial and not semiclassical, in most cases. Semiclassical cases include $N_f=0$ and the case of  arbitrary $N_f$ when the Coulomb branch is not lifted, e.g., the theory described in Section \ref{case1}, with half the quarks periodic and half antiperiodic. 

Next, we consider massive quarks. The superpotential (\ref{wexact}) becomes:
  \begin{equation} \label{wexact2}
 W = (Y {\rm Pf} {\cal{M}})^{1\over N_f - 1} + \eta Y~ + \tr M {\cal{M}},
 \end{equation}
  If the fundamentals are sufficiently massive (heavier than the $Y$ field), a superpotential should be written only in terms of $Y$. This can be done by integrating out the mesons ${\cal{M}}$ by inserting the solution of their equations of motion  (modulo ${\bar{D}}^2$  terms) into (\ref{wexact2}) and obtaining:
  \begin{equation}
\label{superpotential0}
W = a \;Y +   {b \over Y}~,
\end{equation}
where $a$ and $b$ now parameterize the coefficients the two terms.
This superpotential determines the supersymmetric vacuum of the massive theory: $\langle Y\rangle = \pm \sqrt{b \over a}$.
For our study of the center-symmetry realization,  we would like to understand the implications of (\ref{superpotential0}) in more detail and find out: {\it i.)} The relation between the chiral superfield $Y$ and the UV fields of the 4d gauge theory, especially   the relation between its bosonic component and the Wilson line of the gauge field around $\S^1$, i.e.,~the Polyakov loop, and {\it b.)}
The values of the coefficients $a$ and $b$ in the superpotential.
Knowing the answers  will tell us about center-symmetry along the Coulomb branch and will determine the first two terms of the potential in Eq.~(\ref{zerodiracpotential1}). Including small soft supersymmetry breaking and obtaining the rest of the terms in (\ref{zerodiracpotential1}) is then straightforward.

In the next Section, we perform a self-consistent semiclassical calculation of (\ref{superpotential0}). We start by assuming that the Dirac mass $M$ is large enough so that the fundamental supermultiplet  can be integrated out and the long-distance  dynamics can be described only in terms of a superfield $Y$. Furthermore, we assume that we are sufficiently far along the Coulomb branch so that an abelian semiclassical description is valid. Then the   topological excitations contributing to   the superpotential (\ref{superpotential0}) are the BPS and KK monopoles. The effect of the fundamental fermions is to modify their amplitudes by their quantum fluctuations. The  quantum fluctuations depend on the background holonomy (since, along the Coulomb branch, the fundamental superfield mass depends on the holonomy)  as well as on $M$, $L$, and the boundary conditions around $\S^1$. We shall see that the quantum fluctuations due to the massive quark superfields play an important role in determining the holonomy expectation value, which thus acquires $M$-dependence. In the end of the calculation, for the given value of $M$, we have to check that the resulting holonomy expectation value is  far enough  along the Coulomb branch and that the coupling is sufficiently weak, so that the assumption of calculability made in the beginning of this paragraph  is self consistent. 
Self-consistency of semiclassics is then expected  to provide  a lower bound on $M$. 

As usual,  the superpotential is inferred from a computation of the fermion bilinear term generated by  BPS and KK monopole instantons. This calculation has a long history, see \cite{Davies:2000nw}, but some details relevant for the present study were spelled out only recently (in the ${\cal N}=1$ setup, see \cite{Poppitz:2012nz} and references therein). In particular,  it was shown that the bosonic and fermion fluctuation determinants of the vector multiplet in the monopole-instanton backgrounds  do not cancel. This is despite the fact that these  backgrounds preserve supersymmetry and occurs because of the slow fall-off of the background, leading to a 
non-matching density of states of fermions and bosons.
 The non-canceling one-loop determinants modify the tree-level proportionality relation between $\log Y$ and the holonomy $A_3$. In the pure-SYM case, the conclusion that the vacuum is center symmetric is not affected by the quantum-deformed relation between $\log Y$ and $A_3$ \cite{Poppitz:2012nz}. However, we shall see that  this is not so in the theory with fundamentals.  

\subsection{The  supersymmetric vacuum of massive SQCD on $\R^3 \times \S^1$}
\label{susyvacuum}

As stated above, monopole-instantons in massive SQCD generate a fermion-bilinear 't Hooft vertex. We now simply state the result and refer to Appendices for details of the calculation:

\begin{align}
\label{fermionbilinear}
L_{2 \; ferm.} &=  \; \;  {c \over g^6} \; e^{- {vL \over 2 \pi}\left[{8 \pi^2 \over g^2}  + N_f \log {4 \pi \over LM}\right]  + 3 \log {\Gamma(1 - {Lv\over 2\pi}) \over \Gamma({Lv \over 2 \pi})}   + N_f X_{uL}(vL, ML)} \;e^{i \sigma}\; \bar\lambda \bar\lambda   \nonumber \\
&  \;\; +    \; {c \over g^6} \; e^{- (1 -  {vL \over 2 \pi})\left[{8 \pi^2 \over g^2}  + N_f \log {4 \pi \over LM}\right]  -  3 \log {\Gamma(1 - {Lv\over 2\pi}) \over \Gamma({Lv \over 2 \pi})}   - N_f X_{uL}(vL, ML)} \; e^{-i \sigma} \;\bar\lambda\bar\lambda~ + {\rm c.c.}
\end{align}
The expression  (\ref{fermionbilinear}) is obtained by combining the tree level bare 't Hooft suppression factor, the one-loop determinants of the vector and chiral supermultiplets around the BPS and KK monopole instantons (Pauli-Villars-regulated and  computed in Appendix \ref{funddets}) and the integrations over the  bosonic and adjoint-fermion zero modes. 
The origin of the various factors appearing in (\ref{fermionbilinear}), see Appendix \ref{funddets} for  details, is as follows:
\begin{enumerate}
\item The first term gives the contribution of the BPS monopole-instantons, and the second term---the KK monopole-instantons. The BPS monopole-instantons have action ${vL\over 2 \pi} {8 \pi^2 \over g^2}$, where $vL = \pi$ is the center-symmetric point. The KK monopole-instantons have action $(1-{vL\over 2 \pi}) {8 \pi^2 \over g^2}$. At the center-symmetric point, the BPS and KK monopole-instantons have equal action $S_0 = {4 \pi^2 \over g^2}$.
\item
All divergent terms from the determinants are absorbed in a redefinition of the UV cutoff-scale ($\Lambda_{PV}$) coupling in the 't Hooft suppression factor to $g^2$. From now on $g^2$ denotes the 4d gauge coupling of massless SQCD (with   beta function $6-N_f$) evaluated at the scale $4\pi \over L$. 
\item The $e^{\pm i \sigma}$ factors, with $\sigma$---the dual photon field, represent the fact that the monopole-instantons carry long-range fields.  The overall $g^{-6}$ factor comes from zero-mode normalizations, see \cite{Davies:2000nw}. We omit an inessential overall dimensionless constant  $c$; what matters  is that  it is the same for the two terms and  is given in  \cite{Poppitz:2012nz}. Also note that the fermion fields are taken to have dimension 3/2, as in 4d. 
\item 
 The $3 \log({\Gamma(1 - {Lv\over 2\pi})/\Gamma({Lv\over 2 \pi})})$ factor represents the finite part of the one-loop vector supermultiplet determinant. Notice that it  is not a periodic function of $Lv$ and is valid for $0 < Lv < 2 \pi$. The meaning of its singularities is explained in \cite{Poppitz:2012nz}.
\item The factors proportional to $N_f$ arise from the chiral supermultiplet nonzero mode determinants around the monopole-instantons. In addition to   $N_f \log {4\pi \over LM}$ in the square brackets of each term in (\ref{fermionbilinear}),  they also give rise to the terms  denoted by $X_{uL}(vL,ML)$, appearing in the BPS and KK terms:
\begin{equation}
\label{xu}
X_{uL}(vL, ML) = - {2 \over \pi} \sum\limits_{n=1}^\infty {1 \over n} \sin\left(\frac{nvL}{2}\right) \cos(nuL)K_{0}(LM n)~.
\end{equation}
Here, $uL$ denotes the boundary condition for the chiral supermultiplets: for example, $uL=\pi$ for antiperiodic quarks and  $uL=0$ for periodic quarks. If different flavors have different boundary conditions, the corresponding factors of $X_{uL}$ have to be summed over, possibly including different Dirac masses $M$.
\item
The expression for $X_{uL}(vL,ML)$  in (\ref{xu}) converges fast when $ML$$\ge$${\cal{O}}(1)$ (for smaller $ML$, it requires keeping ${\cal{O}}(1/LM)$ terms in the sum). On the other hand, at small $ML$, $X_{uL}$ has a logarithmic singularity, see Eqn.~(\ref{xu3}). Using (\ref{xu3}), we find that, as $ML\rightarrow 0$,   the finite part of the fundamental supermultiplet determinants in the BPS and KK monopole-instanton backgrounds become:
\begin{align}
\label{index1}
e^{N_f\left({vL \over 2\pi} \log LM + X_{uL}(vL,ML)\right)} &\rightarrow  (ML)^{N_f I_{BPS}^{fund.}(vL,uL)} ~,\\
e^{N_f\left((1- {vL \over 2\pi})\log LM - X_{uL}(vL,ML)\right)} & \rightarrow  (ML)^{N_f(1- I_{BPS}^{fund.}(vL,uL))}~. \nonumber
\end{align}
The exponent is expressed in terms of the index counting the fundamental zero modes in the BPS monopole-instanton as a function of holonomy and real mass, as given in \cite{Poppitz:2008hr}:\footnote{Notice that the index is not defined when either of the terms in the square brackets in (\ref{index2}) are exactly integer (at this point, fundamental zero modes become non-normalizable and jump between different monopole-instantons). Also note that, as we show later, in the $ML \rightarrow 0$ limit semiclassics breaks down in all but one case; thus, (\ref{index1}) should be viewed as a consistency check on the calculation, rather than a physical limit to be taken.}
\begin{equation}
\label{index2}
 I_{BPS}^{fund.}(vL, uL)= 1 + \bigg\lfloor{vL + 2 uL \over 4\pi} \bigg\rfloor+ \bigg\lfloor{vL - 2 uL\over 4 \pi} \bigg\rfloor =   \bigg\lfloor{vL + 2 uL \over 4\pi}  \bigg\rfloor-  \bigg\lfloor{ 2 uL - vL\over 4 \pi} \bigg\rfloor~,\end{equation}
 where $\lfloor x \rfloor$ denotes the largest integer smaller than $x$. 
\end{enumerate}
With all factors now explained, let us study the consequences of (\ref{fermionbilinear}). As usual, this will be done by first writing the superpotential implied by $L_{2 \; ferm.}$. To this end, we first redefine the Coulomb-branch modulus field $v$ by expanding around the center-symmetric value $\pi\over L$:
\begin{equation}
\label{bprime}
{L v \over 2 \pi} = {\theta \over 2 \pi} = {1 \over 2} + {g^2 \over 8 \pi^2} b', ~ ~{\rm equivalently }~~ b' = {8 \pi^2 \over g^2} {\theta - \pi \over 2 \pi},
\end{equation} 
where $\theta$ is the angular separation between the Polyakov loop eigenvalues.
We obtain:
\begin{align}
\label{fermionbilinear1}
L_{2 \;ferm.} &=  {c \over g^6}\;  e^{- {4 \pi^2 \over g^2}  - {N_f\over 2} \log {4 \pi \over LM}}  \nonumber \\
& ~~ \times 
\left[ 
e^{- b' \left(1 + {g^2 N_f \over 8 \pi^2} \log{4 \pi \over LM} \right) + 3 \log {\Gamma({1\over 2} - {g^2 \over 8\pi^2}b') \over \Gamma({1\over 2} + {g^2 \over 8\pi^2}b')}   + N_f X_{uL}(\pi + {g^2 \over 4 \pi} b', ML)} e^{i \sigma} \bar\lambda \bar\lambda \right.\nonumber
 \\
& \left. \qquad \;\; +    \; e^{b' \left(1 + {g^2 N_f \over 8 \pi^2} \log{4 \pi \over LM} \right) - 3 \log {\Gamma({1\over 2} - {g^2 \over 8\pi^2}b') \over \Gamma({1\over 2} + {g^2 \over 8\pi^2}b')}   - N_f X_{uL}(\pi + {g^2 \over 4 \pi} b', ML)}  e^{-i \sigma}
\bar\lambda\bar\lambda \; \right]+ {\rm c.c.}\end{align}
The fermion mass term (\ref{fermionbilinear1}) can be obtained from the  superpotential (\ref{superpotential0}), with $a=b$, and $Y \equiv e^{B}$:
\begin{equation}
\label{superpotential1}
W = A\; ( e^B + e^{-B})~, ~~ A \sim {e^{- S_0}(L M)^{N_f\over 2} \over L^{2} g^{2}}~.
\end{equation}
Once again, we do not give the overall numerical constant, but stress that the coefficients of the two terms are identical, 
with our definition of the superfield $B$ (implied above; also see Eq.~(\ref{duality}) below). To leading order in the coupling, the K\" ahler potential of the $B$ field is canonical\footnote{With corrections that can be determined from the moduli space metric  (\ref{vkin1}) after a duality transformation.}
\begin{equation}
\label{kahler1}
K \simeq {g^2 \over 2 (4 \pi)^2 L} B^\dagger B~,
\end{equation}
leading to the bosonic potential
\begin{equation}
\label{bosonic}
V = K_{B^\dagger,B}^{-1}\left|\frac{\partial W}{\partial B}\right|^2=\frac{2 A}{K_{B^\dagger, B}} (\cosh(2\text{Re}B)-\cos(2\text{Im}B))~,
\end{equation}
where $K_{B^\dagger,B}=\frac{\partial^2K}{\partial B\partial B^\dagger}\simeq\frac{g^2}{2(4\pi)^2 L}$.

 To verify that the superpotential $W$, with the coefficient as given  in (\ref{superpotential1}),  leads to the fermion bilinear (\ref{fermionbilinear1}),   recall also that the fermion component enters the chiral superfield\footnote{For a full component expression, see Appendix A of \cite{Intriligator:2013lca} (notice  that their convention corresponds to compactifying $x_2$, in the notation of \cite{Wess:1992cp}).} $B$ as $\sim {L \over g^2} \theta \sigma^3 \bar\lambda$, where $\lambda$ is the 4d gaugino field   (here, $\theta$ is the superspace coordinate, with the usual four-dimensional notation of \cite{Wess:1992cp}  and  compact direction $x^3$). The imaginary part of the lowest component of $B$ is Im$B\vert = i \sigma$, the dual photon field, while 
its real part is expressed through the  holonomy $b'$ (recall the redefinition (\ref{bprime})) as follows:\footnote{The definition of $B$ employed in (\ref{duality}) corresponds to using the variation of $A_3^3$ away from the center-symmetric value as the lowest component of the real linear superfield (describing the dimensional reduction of the  4d gauge supermultiplet)  dual to $B$.  For our study of center-symmetry realization, we find the definition implied by   Eq.~(\ref{fermionbilinear1}), where the entire superpotential is proportional to $(ML)^{N_f\over 2}$, most convenient. Redefining the origin of $B$ would, of course, shift the minimum away from Re$B=0$, but  not the value of the holonomy. For finding the expectation value of the holonomy in the supersymmetric ground state, the corrections to the moduli space metric (\ref{vkin1}) can be neglected, but the determinant factors  in (\ref{fermionbilinear1}), which enter the duality relation (\ref{duality}), can not.  We will be able (maintaining the validity of semiclassics) to take the $ML\rightarrow 0$ limit  only in one case and will see that in this case this factor has a natural origin, see Section~\ref{case1}.}
\begin{equation}
\label{duality}
{\rm Re} B\vert = - b' \left(1 + {g^2 N_f \over 8 \pi^2} \log{4 \pi \over LM} \right) + 3 \log {\Gamma({1\over 2} - {g^2 \over 8\pi^2}b') \over \Gamma({1\over 2} + {g^2 \over 8\pi^2}b')}   + N_f X_{uL}(\pi + {g^2 \over 4 \pi} b', ML)~.
\end{equation}
Equation (\ref{duality}) is the scalar superpartner of the usual photon-dual photon duality. It  takes into account the one-loop modification of the moduli space metric (i.e., the kinetic term of $v$, or of $b'$), see Appendix A of \cite{Poppitz:2012nz} for a recent discussion.

Since the two terms in  the superpotential (\ref{superpotential1}) have identical coefficients,   the supersymmetric minimum, determined by the extremum of the superpotential, $dW =0$, occurs at $\langle B\vert\rangle=0$. It is clear from (\ref{duality}) that $\langle B\vert\rangle  = 0 $ does not imply $\langle b'\rangle=0$, contrary to  the pure-SYM case. 
The vacuum equation, Re$\langle B\vert \rangle =0$, can be written, using:
\begin{align}
\label{3deffective31}
\theta &\equiv  v L  \;\; \;\;({\rm separation \; between \; Wilson \; loop \; eigenvalues}) \nonumber \\
m & \equiv M L\;\;  ({\rm dimensionless \; Dirac \; quark \; mass}) \\
\alpha & \equiv u L \; \;\;\;(\alpha=0-{\rm periodic \; quarks}; \alpha=\pi-{\rm antiperiodic \; quarks}) \nonumber\end{align}
and introducing
\begin{align}
\label{vac1}
  \delta  &\equiv {\theta - \pi \over 2 \pi}~, ~~ b' = {8 \pi^2 \over g^2} \delta ~, \\
{\rm Re}   B\vert(\delta,m,\alpha) &=-  {8 \pi^2 \over g^2}\; {\delta}  + 3 \log{\Gamma({1\over 2} - {\delta}) \over \Gamma({1\over 2}+ {\delta})}    -  N_f \left({\delta} \log{4\pi \over m} - X_\alpha (\pi + 2\pi \delta, m)\right) \nonumber ~.\end{align}

Now we also   impose the condition that the solution of ${\rm Re}   B\vert = 0$ lies within the region of validity of the weak coupling semiclassical expansion. To this end, we 
 recall (see \cite{Poppitz:2012nz}) that supersymmetry relates the derivatives of the  logarithms of instanton determinants with respect to the holonomy  to  the loop-modified moduli-space metric. The effective 3d coupling 
 $g_{3, eff.}^2$, which determines the kinetic term for the holonomy $v$ (as well as for the photon, not shown below):
\begin{equation}
\label{3deffectivemain}
{1 \over2 g_{3, eff.}^2} \; (\partial_i v)^2~,
\end{equation}
 is given by an expression  (\ref{3deffective2}) that depends on the modulus $v$ (or $\delta$, recall (\ref{3deffective31})) as well as the matter content of the theory:\footnote{See  Appendix \ref{deformation} for a derivation and various checks on the formula for the effective coupling.}
\begin{align}
\label{vac2}
{1 \over L g_{3, eff.}^2} &= {8 \pi^2 \over g^2}+    3 \psi ({1\over 2} + \delta) + 3 \psi({1\over 2} - \delta)+  N_f \left(\log {4\pi \over m} - 2 \pi N_f   {\partial  X_\alpha (\theta, m)\over \partial \theta}\big\vert_{\theta= \pi + 2 \pi \delta}\right)\nonumber \\
&= - {\partial\over \partial \delta} \; {\rm Re} B\vert(\delta,m,\alpha) ~.
\end{align}
The last equality in (\ref{vac2}) explicitly shows the stated relation between instanton determinants and moduli space metric.
In order that  the weak-coupling semiclassical approximation be valid, 
  we demand that each of the last two terms (the loop corrections due to adjoints or to fundamentals) on the r.h.s. of (\ref{vac2}) be much smaller than the first ($8\pi^2\over g^2$, the dimensionless `bare' 3d coupling at the cutoff scale $4 \pi\over L$ of the 3d theory). Thus, the vacuum value of the holonomy $\delta_{min}$ determined by Re$B\vert=0$ 
 should also obey:  \begin{equation}
\label{3dweakcondition}
 { 8 \pi^2 \over g^2} \gg   {\rm max} \left\{ \big\vert 3  \psi({\theta_{min} \over 2 \pi})+ 3 \psi(1 - {\theta_{min} \over 2\pi})\big\vert, \big\vert N_f  (\log 4 \pi -  \Phi(\theta_{min},m,\alpha))\big\vert \right\}~, 
\end{equation} where we denoted for future use:\footnote{$\Phi(\theta,m,\alpha)-\Phi(\theta,\Lambda_{PV}L,\alpha)$ is a sum over (s)quark Kaluza-Klein modes, $\sim \sum_{KK}  m_{KK}^{-1}$, see (\ref{phi2}).}
\begin{equation}
\label{phix}
\Phi(\theta, m, \alpha) \equiv \log m + 2 \pi {\partial  X_\alpha(\theta,m)\over \partial \theta}~.
\end{equation}
  
Now we can finally formulate the minimization/consistency problem as follows. The equation that $\delta_{min}$ should solve is:
\begin{align}
\label{vac3}
{8 \pi^2 \over g^2}\; {\delta_{min}}= 3 \log{\Gamma({1\over 2} - {\delta_{min}}) \over \Gamma({1\over 2}+ {\delta_{min}})}    -  N_f \left({\delta_{min}} \log{4\pi \over m} -  X_\alpha (\pi + 2\pi \delta_{min}, m)\right)~.
\end{align}
On the other hand, 
the semiclassical  consistency condition (\ref{3dweakcondition}) requires that the derivative of the l.h.s.~of  (\ref{vac3}) w.r.t.~$\delta$ be much bigger than the derivative of each of the two terms appearing on the r.h.s. Thus, both the vacuum state and the consistency of the  semiclassical approximation can be investigated graphically, for general values $\alpha$ and $m$, and we do so in several examples below (see Fig.~\ref{fig:antiperiodic}).
At the end of this Section, we 
stress again that we keep $g^2 \equiv g_4^2({4 \pi\over L})$ (equivalently, the four-dimensional strong-coupling scale of the $N_f$-flavor theory, $\Lambda$) and $L$ fixed and let $m$, $\alpha$, and $N_f$ vary.

\subsection{Antiperiodic quarks and  Polyakov loop expectation value}
\label{antiperiodic}

We next consider the case of antiperiodic quark supermultiplets and study the Polyakov loop expectation value as a function of $M$. The implications of these results for the Polyakov loop crossover were discussed in Section \ref{qualitative}.

In general, the relation (\ref{duality}), or the equivalent (\ref{vac3}), is complicated and will be studied below. In the limit where the deviation from center symmetry is small, it is clear from (\ref{duality}) that Re$\langle B\vert\rangle  = 0 $ corresponds to, up to corrections suppressed by $g^2$:
\begin{equation}
\label{bprimenoncenter}
\langle b' \rangle \simeq N_f X_\pi(\pi, ML) = - N_f {2 \over \pi} \sum\limits_{n=1}^\infty {1 \over n} \sin\left(\frac{n \pi}{2}\right) (-1)^n K_{0}(LM n)~,
\end{equation}
where we took the case of antiperiodic quarks.
 Thus, for $LM \ge {\cal{O}}(1)$, the deviation from center-symmetry is seen by keeping the first term in the sum (in the qualitative discussion of Section \ref{qualitative},  this expectation value was called $\delta_2$, see Eqs.~(\ref{zerodiracpotential1}, \ref{potexpansion})):
\begin{equation}
\label{bprimenoncenter1}
\langle b' \rangle \simeq   N_f \sqrt{2 \over \pi} {e^{- M L} \over \sqrt{ML}}~.
\end{equation}
This corresponds to an expectation value of the Polyakov loop:
\begin{equation}
\label{ploop1}
\langle \tr \Omega \rangle \approx {g^2 \over 4 \pi} \langle b' \rangle \simeq  N_f  {g^2  \over (2 \pi)^{3/2}} {e^{- ML} \over \sqrt{ML}}~,\end{equation}
valid  in the heavy quark limit ($LM \ge {\cal{O}}(1)$). See Section~\ref{bionpotentialsection} for a ``string breaking" interpretation of (\ref{ploop1}).

\begin{figure}[h] 
   \centering
   \includegraphics[width=6.2in]{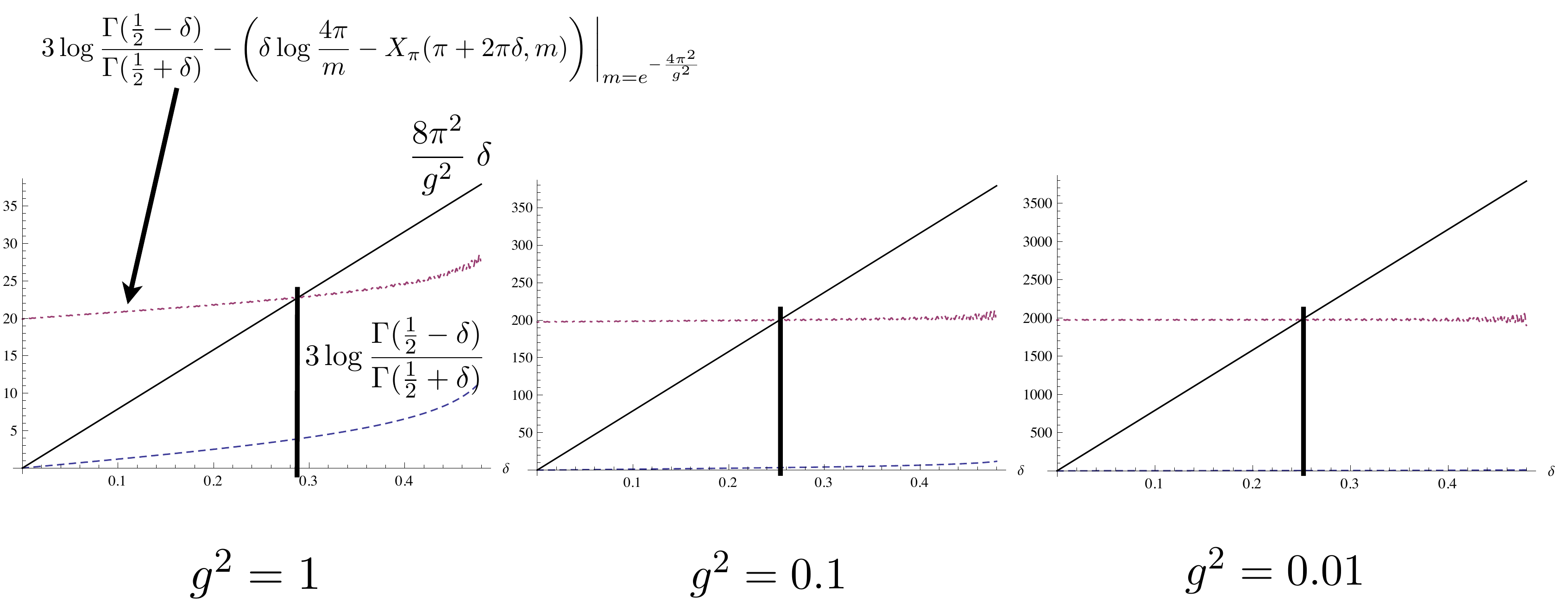} 
   \caption{The graphical determination of $\delta_{min}$ for $M = L^{-1} e^{- S_0}$, for three different values of $g^2$, $N_f=1$, and all antiperiodic quarks, $\alpha=\pi$. This value of $M$ is, within leading exponential accuracy, equal to the dual photon (and superpartners) mass, see Section~\ref{bionpotentialsection}. The three functions of $\delta$  plotted for $0<\delta<.5$ are labeled on the Figure. The relevant intersection point determines that $\delta_{min}  \simeq .25$ for $g^2 \ll 1$. Clearly, the semiclassical approximation is valid, as the slopes of the lines near the intersection points are rather well separated. The plot also shows that even exponentially small values of the Dirac mass  can lead to only finite deviations from center symmetry, e.g., with angular separation of the Polyakov loop eigenvalues about ${\pi \over 2}$, thus still described by a weakly-coupled Abelian theory. }
   \label{fig:antiperiodic}
\end{figure}

For general values of $M$, the  solutions of (\ref{vac3})  can be found graphically---by plotting the function of $\delta$ on the r.h.s.~vs.~the straight line on the l.h.s.~and looking for their intersection---for various values of $g, m, \alpha$. This is illustrated on Fig.~\ref{fig:antiperiodic} for the case when all quarks are antiperiodic ($N_f=1$ is taken on the figure). One conclusion from the Figure is that exponentially small values of the Dirac mass still lead to finite deviations from center symmetry such that the semiclassical Abelian description is valid. We   note that the case shown on Fig.~\ref{fig:antiperiodic} is consistent with our starting assumption: for Dirac masses larger than the dual photon mass one expects  a description that does not involve quark fields.

While we have not determined the precise value of $m=ML$ where the semiclassical description breaks down, clearly this is expected to happen for $M$ sufficiently smaller than the dual photon mass (in principle, this is straightforward to determine numerically). That the breakdown is bound to occur for sufficiently small $M$ follows  from noting that upon a decrease of $m$, the intersection point on Fig.~\ref{fig:antiperiodic} moves to the right---i.e.,~to less separated eigenvalues, toward the point where $SU(2)$ is restored. Simultaneously, the slope of the lines corresponding to the terms on the r.h.s.~of (\ref{vac3}) increases. 

As far as the semiclassical dynamics of topological molecules is concerned, we note that in the ``all-antiperiodic" quark theory, the fundamental zero modes, even for exponentially small masses  (as in Fig.~\ref{fig:antiperiodic}),  have only a short range, $\sim L$, and affect the magnetic and neutral bions  only through their mass insertions and quantum fluctuations.  The  one-loop determinants around monopole-instantons and the related running of the coupling  determine the shift of the vacuum of the theory   away from center-symmetry.

The cases with quarks  that preserve center symmetry, which offer a larger set of tunable parameters, are discussed next.

\subsection{Quarks with real mass  preserving center symmetry}
\label{centersymmetricquarks}

For an $SU(2)$ gauge group, there are  cases, parameterized by a continuous real-mass parameter $\delta u$,  where different fundamental quarks flavors have different boundary conditions, such that the SQCD action has a $\Z_2$ center symmetry which interchanges the two sets of quarks. 
Studying these cases is interesting not because they relate, upon decoupling the superpartners,  to four-dimensional thermal QCD (they do not), but because we hope for additional insight on the effect of fundamental quarks on the  topological excitations as we can vary more parameters. 

To have center symmetry of the SQCD action, consider the theory with $N_f$ flavors, such that $N_f\over 2$ of the flavors have a boundary condition around $\S^1$  given by (\ref{boundary}) with $\alpha_1 = {\pi\over 2} + L \delta u $ and for the other $N_f\over 2$ flavors, given by $\alpha_2 = - {\pi \over 2} + L \delta u$ (in other words, $\alpha_1 - \alpha_2 = \pi$; it suffices to consider $L \delta u \in (0, \pi)$). The center-symmetry transformation, a ``gauge transformation" antiperiodic around the $\S^1$, 
maps the flavors with boundary condition $\alpha_1$ into the flavors with boundary condition $\alpha_2$. In order to be a symmetry of the action, the Dirac masses of the two types of flavors should be identical (notice that no mass term mixing the two types of flavors is allowed).
In SQCD with center symmetry,  
 the superpotential would still be given by (\ref{superpotential1}), but the duality  relation (\ref{duality}) is replaced by: 
 \begin{align}
\label{duality1}
{\rm Re} B \vert &= - b' \left(1 + {g^2 N_f \over 8 \pi^2} \log{4 \pi \over LM} \right) + 3 \log {\Gamma({1\over 2} - {g^2 \over 8\pi^2}b') \over \Gamma({1\over 2} + {g^2 \over 8\pi^2}b')}   \\
& \qquad+ {N_f \over 2} \left[ X_{{\pi\over 2} + L\delta u}(\pi + {g^2 \over 4 \pi} b', ML) + X_{-{\pi\over 2} + L\delta u}(\pi + {g^2 \over 4 \pi} b', ML)\right]~.\nonumber
\end{align}
The effect of the different boundary conditions on $g_{3, eff.}^2$ is, similarly, taken into account by replacing $ X_\alpha$ in (\ref{vac2}) with  ${1\over 2}(X_{\alpha_{1}} + X_{\alpha_{2}})$. 
Now the condition that the expectation value $\langle B\rangle = 0$ is equivalent to $\langle b'\rangle =0$, since at $b'=0$, we have, at the center symmetric point $vL=\pi$:
\begin{align}
\label{centersymmx}
  X_{{\pi\over 2} + L \delta u}(\pi,ML)+& X_{-{\pi\over 2} + L \delta u }(\pi,ML) = \\
  &-\frac{4}{\pi}\sum_{n=1}^\infty\frac{1}{n}\sin\left(\frac{n\pi}{2}\right) \cos({n\pi \over 2}) \cos(L \delta u) K_0(LMn)=0~, \nonumber
 \end{align}
 for any value of the real mass $\delta u$.
Thus, the $\Z_2$ center symmetry of the action is preserved nonperturbatively. This conclusion holds for values of the parameters $M, g, L$, such that the semiclassical description leading to (\ref{duality}) (or (\ref{duality1})) remains valid. 

In the center symmetric vacuum, the physical mass squared of the two types of fundamental flavors are most easily read off the denominator of (\ref{phi2}),  i.e.,
$ {1 \over L^2} \left( {2 \pi k } + {\pi \pm 2 \alpha_i \over 2 }  \right)^2 + M^2$, 
for each type of flavor, $i=1,2$ ($k \in\Z$ is the Kaluza-Klein mode number). Since $2 \alpha_{1,2} = \pm {\pi  } + 2 L \delta u$, the masses are $ {1 \over L^2} \left( {2 \pi k } +   \pi  + L \delta u  \right)^2 + M^2$ and $ {1 \over L^2} \left( {2 \pi k } - L \delta u \right)^2 + M^2$ for $\alpha_1$-flavors and $ {1 \over L^2} \left( {2 \pi k } +    L \delta u   \right)^2 + M^2$ and $ {1 \over L^2} \left( {2 \pi k } + \pi - L \delta u   \right)^2 + M^2$ for $\alpha_2$-flavors. 
Thus,  arbitrarily light massless modes arise, as $M\rightarrow 0$, only if $L \delta u$ is allowed to approach  $0$ (or, equivalently, $\pi$). On the other hand, so long as $L \delta u$ is sufficiently far away from these two values, even at $M=0$, the quark mass in the center symmetric vacuum is  $\sim 1/L$.

 In order not to overly complicate the discussion with general formulae, we shall now consider two limiting cases. In the first ``Case $I.$" we take $L \delta u = {\pi \over 2}$, thus, half the quarks are periodic and half-antiperiodic. As our second ``Case $II.$", we take $L \delta u = 0$, thus allowing for arbitrarily light quarks in the center symmetric vacuum as $M\rightarrow 0$. 

\subsubsection{Case $I.$}
\label{case1}

This is the case of $N_f \over 2$ periodic and $N_f \over 2$ antiperiodic quarks.
As the vacuum is center symmetric,  the theory is Abelian for any $M$. Furthermore, for any value of $M$ the lightest fundamental fermions have mass of order $L^{-1}$ and the unbroken $U(1)$ gauge coupling will be frozen at the scale $\sim {1\over L}$ (this is also easy to check using our expression for the effective coupling (\ref{3deffectivemain})). Thus, we expect now that the dynamics is semiclassically calculable at any arbitrarily small but nonzero $M$, provided $L$ is small enough.
 However, this case is of limited interest for semiclassical bion dynamics as the fundamental fermion zero modes have a very short range $\simeq { 2 L \over \pi}$ and  play no role in bion binding, except for the zero mode mass insertion. Recall that with half the fundamentals periodic and half antiperiodic,  both the BPS and KK monopoles have a fundamental zero mode and thus the overall $M^{N_f \over 2}$ factor in the superpotential is quite natural.
 Thus, as $M\rightarrow 0$, the superpotential vanishes as do all mass scales induced by  $W$. 
As one expects  continuity of the dynamics as a function of holomorphic parameters in supersymmetric theories \cite{Seiberg:1994aj},  this result suggests that the massless theory has a quantum moduli space and that the theory lands on the center-symmetric point when the center-symmetric $M\rightarrow 0$ limit of the $\alpha_{1,2} = (0, \pi)$ fixed-$L$ and fixed-$g^2$ theory is approached. 

To support this picture, we now propose that the superpotential of the massless center-symmetric theory ($N_c=2$ where ${N_f\over 2}$ flavors are periodic and ${N_f\over 2}$ are antiperiodic) for general even $N_f>2$ has the form:\footnote{Notice that this superpotential is different than the one for two different real masses of Ref.~\cite{Aharony:1997bx}, since the points on the Coulomb branch where either periodic or antiperiodic quarks become massless correspond  to enhanced  $SU(2)$ gauge symmetry (as opposed to Case $II.$, see Footnote \ref{ff2}).}\begin{equation}
\label{masslesscase1}
W = (Y {\rm Pf} {\cal{M}}_1)^{2 \over N_f - 2} + (Y^{-1} {\rm Pf}  {\cal{M}}_2)^{2 \over N_f - 2}~,
\end{equation}
and we note that the one remaining  even  value   $N_f=2$ can be similarly considered and that (\ref{masslesscase1}) is valid only away from the origin for $N_f > 4$. Here, 
 ${\cal{M}}_1$ and ${\cal{M}}_2$ are the meson superfields for the two kinds of flavors (${N_f} \times {N_f}$ antisymmetric matrices). 
The center symmetry interchanges ${\cal{M}}_1$ with ${\cal{M}}_2$ and $Y$ with $1/Y$. This superpotential is the sum of the superpotentials for two ${N_f\over 2}$-flavor $SU(2)$ theories away from the origin and does not include any single monopole-instanton terms because they are absent in the present center-symmetric theory, where all monopole-instantons have fundamental zero modes everywhere on the Coulomb branch. When $Y=0$, the ${\cal{M}}_2$-quarks are massive and ${\cal{M}}_1$ are massless (this point corresponds, classically to $v=0$), and v.v.~when $Y=\infty$ (which really stands for the classical point $v = {2 \pi \over L}$, where the second set of quarks becomes massless). 
 
 Our proposed Eqn.~(\ref{masslesscase1}) passes several checks. Integrating out $Y$, one finds $W_{4d} = ({\rm Pf} {\cal{M}}_1{\rm Pf} {\cal{M}}_2)^{1\over N_f- 2}$---the correct superpotential of  4d $SU(2)$ SQCD with $N_f$ flavors away from the origin (it is given by    $ ({\rm Pf} {\cal{M}})^{1\over N_f - 2}$ for $N_c=2$, with ${\rm Pf} {\cal{M}} = {\rm Pf} {\cal{M}}_1 {\rm Pf} {\cal{M}}_2$, i.e.~with  the mesons containing quarks from both kinds of flavors  missing, as they are not part of the effective 3d description (\ref{masslesscase1})). Furthermore, adding  mass terms to (\ref{masslesscase1})  for the two kinds of quarks, $W_{mass} = \tr(M_1 {\cal{M}}_1 + M_2 {\cal{M}}_2)$, and integrating them out, one finds $W \sim {\rm Pf} M_1 Y^{-1} + {\rm Pf} M_2 Y$, i.e. the superpotential of Eqn.~(\ref{superpotential0})  that was computed with great care in the previous Sections. 
Finally, as promised, Eq.~(\ref{masslesscase1}) also  implies that there is a moduli space of vacua. This can be seen, for example, from $Y^2 \sim  {\rm Pf} M_1/{\rm Pf} M_2$ and by noting that any $Y$ can be reached by appropriately adjusting the way the $M_{1,2} \rightarrow 0$ limit is taken. Depending on $N_f$, there are singularities where new massless degrees of freedom appear. When $N_f = 4$, this superpotential should be valid everywhere (the relevant massless mesons and monopole operator are the massless degrees of freedom  \cite{Aharony:1997bx} at vanishing $Y$ or $Y^{-1}$). However, notice that when $M_1 = M_2$ are taken to zero, the theory ends on the center-symmetric point $Y^2 = 1$ on the moduli space, consistent with the result of our semiclassical analysis.
  
\subsubsection{Case $II.$}
\label{case2}

This is the case where $N_f \over 2$ flavors have boundary condition (\ref{boundary}) given by $\alpha_1 = {\pi \over 2}$ and the other $N_f \over 2$ flavors -- by $\alpha_2 = - {\pi \over 2}$. The vacuum is center symmetric, see (\ref{centersymmx}), and the dynamics is, once again, Abelian. However, for $vL = \pi$, and  $\alpha_{1,2} =   \pm {\pi \over 2}$, i.e. with $L \delta u=0$, there is    a charged mode of  mass determined solely by $M$, an external tunable parameter. We thus expect that as $M$ is decreased, the quantum modification of the moduli-space metric will take over the tree-level value,\footnote{\label{ff2}A   dual description of the massless limit of  Case $II.$ with $N_f=2$ was proposed  in \cite{Aharony:1997bx}.} leading to a breakdown of semiclassics. Nonetheless, it is of interest to study this case, as it  gives us an arena to study the effect of (hopefully, arbitrarily) light fundamental zero modes on the weak-coupling semiclassical bion dynamics. At small $ML$, the contribution to the 3d effective coupling can be inferred either from the Poisson re-summed expression (\ref{phi1}) for $\Phi(\pi, m, {\pi\over2})$, by numerically plotting the sum (note, however, that this  requires including an increasing number, of order $1/m$,  of terms in the sum (\ref{phi1}); doing so numerically reproduces the result given below). It is simpler to start from the Kaluza-Klein sum expression (\ref{phi2}), which shows that ($m=ML$):\begin{equation}
\label{phicase2}
\Phi(\pi, m, \pm{\pi \over 2}) \simeq - {\pi \over 2 m} ~ {\rm as } \; m \rightarrow 0~.
\end{equation}
Numerically, this asymptotic expression is already valid for $m \sim 0.1$.
Then, recalling the condition (\ref{3dweakcondition}),  we conclude from (\ref{phicase2}) that the weak-coupling expansion for Case $II.$ is valid whenever ${8 \pi^2 \over g^2} \gg {N_f \pi \over m}$, or $M  \gg  {1 \over 8 \pi N_f}{g^2 \over L}$. 
Thus the range of the fundamental-fermion zero mode would be of order ${1\over M} \ll  {8 \pi N_f  L \over g^2} = 2 N_f {4 \pi L \over g^2} = 2 N_f r_{\rm b}$ (here $r_{\rm b}$ is the  bion size (\ref{bionsize})).  Thus, one would be tempted to conclude that semiclassics might allow, at best, some marginal role for fundamental-fermion exchange in the bion formation---especially at large $N_f$ (which is at most 5 for $SU(2)$) in the  window $r_{\rm b} \ll {1\over M} \ll 2 N_f r_{\rm b}$.
In the following Section we argue that there is no contribution to binding from exchange of fundamental fermions in the bion action that follows from the potential (\ref{bosonic}).

\subsection{The bion-induced potential}
\label{bionpotentialsection}

Consider now the scalar potential following from (\ref{superpotential1}) and (\ref{kahler1}).
In our convention of (\ref{superpotential1}, \ref{fermionbilinear1}), the first term in $W$ is due to BPS monopoles and the second---due to KK monopoles. Recall that the fundamental zero modes (for zero $M$) for periodic quark superfields are localized on the BPS monopoles (and for antiperiodic fields---on the KK monopoles). 
 The scalar potential that follows from $W$, $K$, omitting loop corrections to the K\" ahler potential (which multiply the entire potential) is:
\begin{equation}
\label{pot1}
 V \sim {L \over g^2}\; A^2 \; |W'|^2 = {(LM)^{N_f} e^{- {8 \pi^2 \over g^2}} \over L^3 g^6}  \left( e^{ 2 {\rm Re}B\vert} +   e^{ - 2 {\rm Re}B\vert} - e^{2 i  {\rm Im} B\vert } - e^{ - 2 i {\rm Im} B\vert} \right) ~,
\end{equation}
where we repeat Eq.~(\ref{duality}) for convenience: 
\begin{align}
\label{duality2}
{\rm Re} B\vert &= - b' \left(1 + {g^2 N_f \over 8 \pi^2} \log{4 \pi \over LM} \right) + 3 \log {\Gamma({1\over 2} - {g^2 \over 8\pi^2}b') \over \Gamma({1\over 2} + {g^2 \over 8\pi^2}b')}   + N_f X_{uL}(\pi + {g^2 \over 4 \pi} b', ML)~, \nonumber \\
{\rm Im} B\vert & = \sigma~,
\end{align}
and, if different flavors have different boundary conditions ($uL$) a corresponding sum is understood in the first line above. Recall also that $g^2 b' \over 8 \pi^2$ is the deviation of $Lv \over 2 \pi$ from the center-symmetric value $1\over 2$, see (\ref{bprime}). Finally, we repeat  the expression for $X_{uL}$ (\ref{xu}): 
\begin{equation}
\label{xu4}
X_{uL}(vL, ML) = - {2 \over \pi} \sum\limits_{n=1}^\infty {1 \over n} \sin\left(\frac{nvL}{2}\right) \cos(nuL)K_{0}(LM n)~.
\end{equation}
While this sum converges fast when $ML > {\cal{O}}(1)$, at small $ML$, $X_{uL}$ has a logarithmic singularity, and, as we discussed in (\ref{index1}, \ref{index2}) 
it is now clear that this introduces an asymmetry between BPS-anti-BPS and KK-anti-KK neutral bions. Note that (\ref{index1}) implies that
$e^{\pm 2 N_f X_\pi(\pi, ML)} \approx (ML)^{\mp N_f}$ and $e^{ \pm 2 N_f X_0(\pi, ML)} \approx (ML)^{\pm {N_f}}$ for small $ML$.
Thus,  in the approximately center-symmetric vacuum (\ref{bprimenoncenter1}) of the all-antiperiodic quark theory, the overall $(LM)^{N_f}$ factor in the scalar potential (\ref{pot1}) cancels in the first (BPS-anti-BPS bion) and squares in the second  (KK-anti-KK bion) term in the scalar potential---consistent with the absence (presence) of fundamental zero modes in the BPS (KK) monopoles for antiperiodic quarks.

On the other hand, recall Section~\ref{case1}, in Case $I.$, where the sum $X_0+ X_\pi$ appears in (\ref{duality2}), we find that 
the overall $M^{N_f}$ factor in (\ref{pot1}) is physical and corresponds to the fact that both the BPS and KK monopoles carry zero modes (see (\ref{xucenterlimit}) for an expression relevant for Case $II.$).

  As already stated, the first terms in $V$ are due to BPS-anti-BPS ``center stabilizing" neutral bions, the second---due to KK-anti-KK neutral bions, and the last two are due to magnetic bions. 
    In the case of all antiperiodic quarks, written in terms of the   deviation from center-symmetric holonomy, $b'$, and the dual photon (${\rm Im} B= \sigma$) instead, the potential looks like:
  \begin{equation}
  \label{allantipotential}
  V \sim e^{ - 2 b' + 2 N_f X_\pi(\pi, ML)} + e^{2 b' - 2 N_f X_\pi(\pi, ML)} - \cos{2 \sigma}~,
  \end{equation} where  we assumed that the deviation from center symmetry is small, as in (\ref{bprimenoncenter}). Comparing with our conjectured potential of Eq.~(\ref{zerodiracpotential1}), it is clear that the ratio of the KK-anti-KK neutral bion amplitude to that of the BPS-anti-BPS bion amplitude is as described there by $\delta_2$ (with  $ {\delta_1=2\delta_2= 2 N_f X_\pi(\pi,ML)}$)  and that, see (\ref{bprimenoncenter}, \ref{bprimenoncenter1}), at large mass (and fixed $L$), the KK-anti-KK amplitude is suppressed (or the BPS-anti-BPS one is enhanced). This is also the behavior expected at small $M$, when one expects a suppression of the KK-anti-KK neutral bions due to the  mass insertion suppression. Notice, however, that as we increase $L$ at fixed $M$ the center symmetry restores exponentially. We interpret this as the diminishing influence of the winding matter modes which screen the Polyakov loop (i.e., heavy quark), which become suppressed as the radius $L$ is increased (the Poisson resummed expression for $X$, Eq.~(\ref{xu4}) has an interpretation as a sum over worldlines of massive particles winding around the $\S^1$).

 Next, turning to the case with light fermions in the center symmetric vacuum, Case $II.$, we note that 
the contributions of fermions to the bion action cannot be interpreted as the binding due to fundamental zeromode exchange, for the following reason: the amplitudes of the fundamental exchange must necessarily be dependent on the effective masses of the fundamental fermions (and therefore on their range), i.e., ${v_{min}\over 2}\pm u $, and therefore on the real mass $u$. However it is clear that in the symmetric situation when we take $N_f\over 2$ flavors to have $u={\pi\over 2L}+\delta u$ and $N_f\over 2$ flavors to have $-{\pi \over 2L}+\delta u$  by equation (\ref{centersymmx}) and since all the dependence on the real mass $\delta u$ is in $X_{\pm{\pi \over 2} + L \delta u}(vL,ML)$, the neutral bion actions in \eqref{pot1} are, in fact, $\delta u$ independent, which contradicts the assumption that fundamental zeromode exchange contributes to the neutral bions. We therefore think that the fundamental zeromode exchange in the dilute semiclassical regime does not contribute to the bosonic potential. 
 
 In this regard, we note  that a study of the ``streamline" configuration for instanton-anti-instanton molecules on $\R^3 \times \S^1$, along the lines of the related $\R^4$ study of SQCD of Ref.~\cite{Yung:1987zp}, and of its relation to the Bogomolnyi--Zinn-Justin analytic continuation is an interesting problem for the future. We believe that such a study will indeed reveal that the fundamental exchange does not contribute to the neutral bion binding.\footnote{A related question, already posed in \cite{Poppitz:2011wy}, that a study of the streamline for monopole-instantons might help answer is that of the absence of magnetic bions in ${\cal N}=2$ Seiberg-Witten theory on $\R^3 \times \S^1$. Only a supersymmetry argument is known to date but a dynamical understanding would  be desirable; see  Section 8.1 in \cite{Argyres:2012ka} for  additional motivation.} 
 In   this paper, we only studied the supersymmetry-based derivation of the potential (\ref{zerodiracpotential1}).

Our final remark is that it is easy to see that for the supersymmetric theory with massive matter multiplet the correlator of the two Polyakov loops exhibits the string breaking at some distance. Computing the correlator
\begin{equation}
\label{loopcorrelator}
\avg{\tr \Omega (x) \tr \Omega^\dagger(y)}=\avg{\cos(\frac{v(x)L}{2})\cos(\frac{v(y)L}{2})}\approx \left(N_f \frac{g^2}{(2\pi)^{3/2}}\right)^2\frac{e^{-2ML}}{ML}+ \frac{g^2}{4\pi}    \frac{e^{-m_{el}r}}{rL}~,
\end{equation}
where we have written $v(x)=v_0+\delta v(x)$, where $v_0 \simeq \pi \pm {g^2 \over 4 \pi L} \avg{b'}$ is the vev of $v(x)$, see (\ref{bprime}, \ref{bprimenoncenter1}), and using the $v$-field correlator $\avg{\delta v(x)\delta v(y)}\approx {g^2 \over 4 \pi L} {e^{-m_{el} r}\over  r}$, where $m_{el}$ is the $b'$ (and dual photon $\sigma$) mass. The interpretation of the above result is clear: the first term is very small for large masses and the second term dominates in the sum for $m_{el}r<2ML$. However when the term $m_{el}r$ (the energy of the string) becomes of the order of $2ML$ (the energy to produce a quark antiquark pair) the first term dominates  and the free energy of a quark-antiquark is simply the energy to produce a pair that will screen the potential. 

\section{Conclusions}
\label{conclusion}
Our main result is  that the observation of Ref.~\cite{Unsal:2010qh,Poppitz:2012nz, Poppitz:2012sw}---where a  semiclassically calculable center-symmetry breaking transition in softly-broken SYM on $\R^3\times S^1$ was found and argued to be  continuously connected, upon decoupling the superpartners, to the thermal deconfinement transition in pure YM theory---can only be partly replicated in more realistic situations involving fundamental quarks.

Specifically, we argued that the semiclassical picture of the transition, driven by the competition between perturbative fluctuations and nonperturbative topological ``center-stabilizing bions" and monopole-instantons holds in softly broken SQCD, albeit only for quarks of mass larger than (roughly) the dual photon mass, i.e., the confinement scale in the long-distance theory on $\R^3 \times \S^1$. 
For such sufficiently heavy quarks, we showed that a crossover behavior of the Polyakov loop from an approximately center-symmetric to center-broken occurs as one increases the ``temperature" $1\over L$, as observed in thermal lattice QCD with massive quarks. The dynamical reason for  the crossover behavior---the deformation of the constituent monopole-instantons by the quarks' quantum fluctuations---is qualitatively explained in Section \ref{qualitative} and    quantitatively studied in Section \ref{susyvacuum}.

For lighter quarks, the calculable semiclassical picture of the vacuum and of the Polyakov-loop crossover fails and a dual description becomes necessary. Such dual descriptions of massless SQCD on $\R^3 \times \S^1$ have been proposed and studied in some detail, see   \cite{Aharony:2013dha} and references therein. Their microscopic origin and any possible connection to deconfinement remain interesting subjects for future studies.

Finally, we speculate what happens in QCD with light quark flavors. Since in QCD chiral symmetry is broken at sufficiently low temperatures, the fermions always have an effective mass. On the other hand, the dilute gas approximation is invalidated as the coupling is strong. However in the studies of models of the vacuum---the instanton-liquid  (see \cite{Schafer:1996wv} and references therein) and more recently the instanton-monopole liquid \cite{Shuryak:2012aa,Faccioli:2013ja}---it is precisely the high density of topological defects that was seen as contributing to the formation of the chiral condensate and the onset of chiral symmetry breaking. The contribution to the vacuum in that case should not be seen as isolated topological objects (or pairs thereof), but of a network of highly correlated---by the exchange of fundamental fermions---objects. 

 \acknowledgments

We thank Mithat \" Unsal for comments.
E.P.~was supported in part by the National Science 
and Engineering Council of Canada (NSERC). T.S. is supported by BayEFG.

\appendix

\section{Taming the perturbative contributions to the Polyakov loop potential}
\label{taming}

Here, we address the immediate worry   that fundamentals will destabilize the center symmetry of $\R^3 \times \S^1$. However, since the boundary conditions on $\S^1$ are supersymmetry-preserving, in the supersymmetric limit the  holonomy potential vanishes. When a small gaugino mass $m_{soft}$ is introduced, as in \cite{Poppitz:2012sw}, it generates a one-loop soft scalar mass for the fundamental fields---but not to the fermions, as perturbative contributions to their mass are protected by the classical $U(N_f)_L \times U(N_f)_R$ symmetry. The one-loop soft scalar mass squared scales as $m_0^2 \sim {g^2\over 4 \pi^2} m_{soft}^2$.\footnote{A precise expression can be borrowed from  the literature on ``gaugino-mediation" of supersymmetry breaking, but we will not need it (in fact, $m_0$ should be treated as an independent small supersymmetry-breaking parameter whose order of magnitude is as given).} The one loop potential  \cite{Gross:1980br}  due to $N_f$ fundamental Dirac fermions of mass $M$ and their scalar superpartners of mass $M_0$, $M_0^2 \equiv M^2 + m_0^2$, can be easily shown to read:\footnote{Here, we consider the one-loop potential for general $N_c$; in addition, note that for $N_c=2$, $v_1 $and $v_2$  in (\ref{vev1}) have a different normalization from the one chosen in Eq.~(\ref{higgs}).} 
\begin{equation}
\label{GPYfundamentals}
V(v_1, ... v_{N_c}) = {2 \over \pi^2} \sum\limits_{a = 1}^{N_c} \sum\limits_{p=1}^\infty {\cos p v_a \over p^2} \left( K_2 (p M) M^2 - K_2 (p M_0) M_0^2\right) ~,
\end{equation} 
where $v_a$ and all mass scales are given in units of $L$, the size of $\S^1$, which is set to unity until further notice.
The potential $V$ is a function of $\{v_a\}$, the eigenvalues of the holonomy around $\S^1$, more precisely defined as:
\begin{equation}\label{vev1}
\sum\limits_{a = 1}^{N_c-1} A_3^a T^a  + B_3 {\bf 1}= {\rm diag}(v_1, v_2, \ldots v_{N_c})~,
\end{equation}
where $T^a$ ($\tr T^a T^b = \delta^{ab}/2$) are the $SU(N_c)$ gauge group generators in the fundamental representation, $\bf 1$ denotes the $N_c \times N_c$ unit matrix, and $B_3$ is the $\S^1$ component of a non-dynamical $U(1)$ background gauge field (it can be thought  of as ``gauging" the vector like baryon number symmetry). The expectation value of $B_3$ is usually called ``real mass" in the supersymmetric literature as it provides a real chirally-symmetric mass term in the long-distance 3d theory. In order to not generate a Chern-Simons term in the $\R^3 \times \S^1$, $B_\mu$ would have to correspond to an anomaly-free global symmetry in 4d and induce no parity anomaly in 3d (the relevant conditions are formulated in \cite{Poppitz:2008hr}).
 The holonomy corresponding to (\ref{vev1}) is:
\begin{equation}
\label{holonomy}
\Omega = {\rm diag}\left( e^{i v_1}, e^{i v_2}, \ldots, e^{i v_{N_c}}\right)~.
\end{equation}
With $B_3=0$, the center-symmetric state corresponds to ${v_k}$ such that $\sum\limits_{k=1}^{N_c} e^{i v_k} =0$ (so that $\tr \Omega = 0$) and $\sum\limits_{k=1}^{N_c} v_k = 0\;{\rm mod}\; (2 \pi)$ (ensuring $\det \Omega = 1$).

When the supersymmetric Dirac mass $M \rightarrow 0$, the relation $K_2 (p M)  M^2 =  {2\over p^2} + {\cal{O}}(M^2)$ shows that the first term in (\ref{GPYfundamentals}) reduces to the well-known expression for massless fundamental fermions, favoring a center-broken vacuum (e.g., for $N_c = 2$, $v_1 = - v_2 = \pi$). In the limit when the supersymmetry-breaking soft mass $m_0^2 \ll M^2$, we have instead, to leading order in $m_0^2$ (recall that $L=1$ in all expressions here):
\begin{equation}
\label{GPYfundamentals2}
V(v_1, ... v_{N_c}) = m_0^2 \; {2 \over \pi^2} \sum\limits_{a = 1}^{N_c} \sum\limits_{p=1}^\infty {\cos p v_a \over p^2} \left( {M p \over 2} K_3(M p) - 2 K_2(M p)\right)~.
\end{equation} 
It is easy to see that the potential (\ref{GPYfundamentals2}) still prefers center-broken holonomy around the $\S^1$; this remains true also in the case the supersymmetric Dirac mass $M$ vanishes.  Just as in the pure SYM case of \cite{Poppitz:2012sw}, the perturbative effect of the fundamental flavors
 is suppressed by the small supersymmetry-breaking parameter $m_0^2$. This parameter is at our disposal and will be taken exponentially small so that the perturbative effects in (\ref{GPYfundamentals2}) can be subleading with respect to the interesting nonperturbative objects: monopoles, magnetic- and center-stabilizing bions.

\section{Fundamental determinants in monopole-instanton backgrounds}
 \label{funddets}
 
For a single Dirac  fundamental matter supermultiplet, the one loop determinants around a BPS or KK monopole instanton contribute
\begin{equation}
\frac{\sqrt{\det{(\Delta_-+M^2)(\Delta_++M^2)}}}{\det{(\Delta_++M^2)}}\;,
\end{equation}
where, for a self-dual background,
\begin{equation}
\Delta_-=-D_\mu^2 + 2 \; {\bm{ \sigma \cdot B}} \;,\quad \Delta_+=-D_\mu^2\;,
\end{equation}
where $\bm B$ is the magnetic field of the relevant monopole-instanton and we are using the notation of \cite{Poppitz:2012nz}.
The determinants in the numerator are the fermionic contribution ${\rm det}(\slashed D - M)$ and in the denominator---the contributions of the two complex fundamental scalars (see, e.g., \cite{Vandoren:2008xg} for an introduction). The Pauli-Villars renormalized contribution is then
\begin{equation}
\sqrt{\frac{\det{(\Delta_-+M^2)}}{\det(\Delta_++M^2)}\frac{\det{(\Delta_++\Lambda_{PV}^2 +M^2})}{\det(\Delta_-+\Lambda_{PV}^2+M^2)}}~.
\end{equation}
Now, defining the index function (note that, in 4d self-dual backgrounds, $I_F(\mu^2)$ is actually $\mu$-independent and is equal to the index at any $\mu$; this is, however, not so in a 3d monopole-instanton background, where the index is recovered at $\mu=0$ only):
\begin{equation}
I_F(\mu^2)=\tr\frac{\mu^2}{\Delta_-+\mu^2}-\tr\frac{\mu^2}{\Delta_++\mu^2}~,
\end{equation}
and using the relation
\begin{equation}
\int_{\Lambda_{PV}^2+M^2}^{M^2}\frac{d\mu^2}{\mu^2} I_F(\mu^2) =\ln\left(\frac{\Delta_-+M^2}{\Delta_-+\Lambda_{PV}^2+M^2}\right)-\ln \left(\frac{\Delta_++M^2}{\Delta_++\Lambda_{PV}^2+M^2}\right)~,
\end{equation}
we have that, for a single Dirac fundamental flavor:
\begin{equation}
\label{diracflavordet}
\sqrt{\frac{\det{(\Delta_-+M^2)}}{\det(\Delta_++M^2)}\frac{\det{(\Delta_++\Lambda_{PV}^2+M^2})}{\det(\Delta_-+\Lambda_{PV}^2+M^2)}}=\exp\left[{\frac{1}{2}\int_{\Lambda_{PV}^2+M^2}^{M^2}\frac{d\mu^2}{\mu^2}}I_{F}(\mu^2)\right]~.
\end{equation}
The index function $I_F(\mu^2)$ for fundamental fermions in a BPS instanton-monopole background was calculated in \cite{Poppitz:2008hr} with the result:
\begin{equation}
\label{indexbps}
I_{F}^{BPS}(\mu^2)=\frac{Lv}{2\pi}+\frac{1}{2}\left[\sum_{n=-\infty}^\infty\frac{\frac{2\pi n}{L}+\frac{v}{2}}{\sqrt{\left(\frac{2\pi n}{L}+\frac{v}{2}\right)^2+\mu^2}}-(v\rightarrow-v)\right]\;~.
\end{equation}
Similarly, for KK instanton-monopoles, it was found that:
\begin{equation}
\label{indexkk}
I_{F}^{KK}(\mu^2)=1 - \frac{Lv}{2\pi} - \frac{1}{2}\left[\sum_{n=-\infty}^\infty\frac{\frac{2\pi n}{L}+\frac{v}{2}}{\sqrt{\left(\frac{2\pi n}{L}+\frac{v}{2}\right)^2+\mu^2}}-(v\rightarrow-v)\right]\;~.
\end{equation}
We note that $v$ in the above equations  is related to the parameters in Eq.~(\ref{vev1}) as: $v = 2 v_1 = - 2 v_2$ (e.g., the center-symmetric point is $v=\pi/L$; recall $L=1$ in (\ref{vev1})). 
The sum  in the above integrals is formally divergent and we regularize it by analytic continuation. First, we introduce the sum:
\begin{equation}
\label{fdef}
F(s;a,c)=\sum_{n=-\infty}^{\infty}\frac{1}{\left[(n+a)^2+c^2\right]^s}~,
\end{equation}
for which it is  well-known that (see, e.g., the Appendix of \cite{Ponton:2001hq} for a derivation):
\begin{equation}
\label{fexpr}
F(s;a,c)=\frac{\sqrt{\pi}}{\Gamma(s)}|c|^{1-2s}\left(\Gamma\left(s-\frac{1}{2}\right)+4\sum_{n=1}^\infty (\pi n |c|)^{s-1/2}\cos(2\pi na)K_{s-1/2}(2\pi n|c|)\right)~.
\end{equation}
Then, we note that the sum we need to evaluate (from  the square brackets of (\ref{indexbps}) and (\ref{indexkk})) can be expressed in terms of $F(s,a,c)$: 
\begin{equation}
S(a,c)= \sum_{-\infty}^\infty \frac{n+a}{\sqrt{(n+a)^2+c^2}}= \partial_a F(-{1\over 2};a,c)~,
\end{equation}
where we introduced $a=vL/(4\pi)$ and $c=\mu L/(2\pi)$. Therefore, from (\ref{fexpr}) we have:\footnote{As a check, if we take $\mu\rightarrow 0$ (i.e., $c\rightarrow 0$) we obtain:
\begin{align}\nonumber
\lim_{c\rightarrow 0}S(a,c) =\frac{2}{\pi}\sum_{n=1}^\infty \frac{1}{n}\sin(2\pi n a)=\frac{i}{\pi}\left[\ln(1-e^{2\pi i a})-\ln(1-e^{-2\pi i a})\right]= \frac{i}{\pi}\ln\left(-e^{2\pi i a}\right)=1-2 (a-\lfloor a\rfloor)~,
\end{align}
where $\lfloor a \rfloor$ denotes the largest integer smaller than $a$.
Thus the index (\ref{indexbps}) becomes  (c.f.~Eq.~(3.5) of \cite{Poppitz:2008hr}):
\begin{equation}
\nonumber
I_{F}^{BPS}(0)={L v \over 2 \pi} + {1 \over 2} (S(a,0) - S(-a,0))=  \left\lfloor \frac{vL}{4\pi}\right\rfloor-\left\lfloor\frac{-vL}{4\pi}\right\rfloor~ = 1 - I_F^{KK}(0)~.
\end{equation}}
\begin{equation}
\label{s2}
S(a,c)=4|c|\sum_{n=1}^\infty \sin(2\pi n a)K_{-1}(2\pi n |c|)~.
\end{equation}
The general expression for the index function,  substituting (\ref{s2}) into (\ref{indexbps}, \ref{indexkk}),  is:
\begin{equation}
I_F^{BPS}(\mu^2)=\frac{vL}{2\pi}+\frac{2\mu L}{\pi}\sum_{n=1}^\infty \sin\left( \frac{nvL}{2}\right)K_{1}(n\mu L)
\end{equation}
for the BPS monopole-instanton, and, for the KK monopole: 
\begin{equation}
I_F^{KK}(\mu^2)=1- \frac{vL}{2\pi} - \frac{2\mu L}{\pi}\sum_{n=1}^\infty \sin\left( \frac{nvL}{2}\right)K_{1}(n\mu L)~.		
\end{equation}

For future use, we note that the above equations for $I_F(\mu^2)$ are easily generalized by introducing a so-called ``real mass'' to the fundamental supermultiplet, i.e., an overall $U(1)$ shift to $A_3$ of Eq.~(\ref{vev1}), denoted there by $B_3$. Taking $B_3  = u$, we obtain:\footnote{By replacing $v/2 \rightarrow v/2 + u$ in the first sum, and $- v/2 \rightarrow - v/2 + u$ in the second sum  in the square brackets  in (\ref{indexbps}) and (\ref{indexkk}). We also note that the $U(1)$ symmetry that $B_3$ gauges is vectorlike and no Chern-Simons term is induced in the long-distance theory on $\R^3 \times \S^1$ \cite{Poppitz:2008hr}.}\begin{align}
\label{realmassindexfunctions}
I_F^{BPS}(\mu^2)&=\frac{vL}{2\pi}+\frac{2\mu L}{\pi}\sum_{n=1}^\infty \sin\left( \frac{nvL}{2}\right)\cos(nuL)K_{1}(n\mu L) ~,\nonumber \\
 I_F^{KK}(\mu^2)&=1- \frac{vL}{2\pi} - \frac{2\mu L}{\pi}\sum_{n=1}^\infty \sin\left( \frac{nvL}{2}\right)\cos(nuL)K_{1}(n\mu L)~.
\end{align}\

Now we continue to evaluate the determinants with $u=0$. For $N_f$ fundamental Dirac flavors, we find,  for the BPS monopole instanton background:\footnote{At this stage, we replaced $\sqrt{\Lambda_{PV}^2+M^2}$ by $\Lambda_{PV}$, which is valid as long as we do not demand that our formulae  automatically describe the  decoupling limit of heavy flavors.}
\begin{align}
\label{fundbpsdeterminant}
{N_f \over 2} \int_{\Lambda_{PV}^2}^{M^2} \frac{d\mu^2}{\mu^2}I_F^{BPS}(\mu^2)&=N_f \frac{vL}{2 \pi}\ln\left(\frac{M} {\Lambda_{PV}}\right)+N_f \frac{2  L }{\pi}\sum_{n=1}^\infty \sin\left(\frac{nvL}{2}\right)\int_{\Lambda_{PV}}^{M} d\mu \;K_{1}(nL\mu)\nonumber\\
&=N_f \frac{vL}{2 \pi}\ln\left(\frac{M}{\Lambda_{PV}}\right)+N_f \frac{2}{\pi }\sum_{n=1}^\infty \frac{1}{n}\sin\left(\frac{nvL}{2}\right)\left(K_0(L\Lambda_{PV}n)-K_{0}(LM n)\right)\nonumber\\
&=N_f \frac{vL}{2 \pi}\ln\left(\frac{M}{\Lambda_{PV}}\right)- N_f\frac{2}{\pi }\sum_{n=1}^\infty \frac{1}{n}\sin\left(\frac{nvL}{2}\right)K_{0}(LM n) ~,
\end{align}
while for the KK monopole instanton, we similarly obtain:\begin{equation}\label{fundkkdeterminant}
{N_f \over 2} \int_{\Lambda_{PV}^2}^{M^2} \frac{d\mu^2}{\mu^2}I_F^{KK}(\mu^2) = N_f (1 - \frac{vL}{2 \pi}) \ln\left(\frac{M}{\Lambda_{PV}}\right)+N_f \frac{2}{\pi }\sum_{n=1}^\infty \frac{1}{n}\sin\left(\frac{nvL}{2}\right)K_{0}(LM n) ~.\end{equation}

For completeness, we also give the expression for the nonzero mode determinant of the vector supermultiplet from Ref.~\cite{Poppitz:2012nz}:
\begin{equation}
\label{dets1}
\Delta_{adj} \equiv \left( { \det \Delta_+ \over \det^\prime \Delta_- } \;  {\det \Delta_- + \Lambda_{PV}^2 \over \det \Delta_+ + \Lambda_{PV}^2 } \right)^{3\over 4}\bigg\vert_{BPS} = \left({4\pi \over L}\right)^3 e^{6 {L v \over 2 \pi} \log {\Lambda_{PV} L \over 4 \pi} + 3 \log {\Gamma(1 - {Lv\over 2 \pi}) \over \Gamma({L v \over 2 \pi})}}~,
\end{equation}
while the corresponding expression for the KK monopole-instanton is obtained by substituting ${vL\over 2\pi} \rightarrow 1 - {vL\over 2\pi}$. We note that the vector supermultiplet determinant is dimensionful, since the zero modes of $\Delta_-$, as indicated by the prime in (\ref{dets1}), are excluded from the ratio of determinants. 

\section{Monopole-instanton determinants and effective coupling}
\label{couplingappx}

The derivatives of the logarithm of the one-loop nonzero mode determinants around the monopole-instantons w.r.t. the modulus field $v$
  determine the moduli space metric  and, thus, the effective coupling of the long-distance 3d theory. The relation between the one-loop determinants and the one-loop moduli space metric is implied by supersymmetry and is explained in \cite{Poppitz:2012nz}. 
 Explicitly, 
the modulus field $v$ has a kinetic term ($i=1,2,3$) given by the sum of the bare coupling and the one-loop contributions of the adjoint and fundamental supermultiplets
\begin{equation}
\label{vkin1}
{1 \over 2} (\partial_i v)^2\left[ {L \over g_4^2(\Lambda_{PV})} - {1\over 4 \pi} F_{adj}(v,L) - {1\over 4 \pi} F_{fund}(v,L,M,u)\right] \equiv {1 \over2 g_{3, eff.}^2} (\partial_i v)^2~,
\end{equation}
where $g_4^2(\Lambda_{PV})$ is the 4d bare gauge coupling.

The derivative of the fundamental BPS determinant (\ref{fundbpsdeterminant}) (taken with $u \ne 0$ below) w.r.t. $v$ reads:
\begin{align}
\label{fund1}
F_{fund} (v, M, L, u) & \equiv {d \over d v}\left[
{N_f \over 2} \int_{\Lambda_{PV}^2}^{M^2} \frac{d\mu^2}{\mu^2}I_F^{BPS}(\mu^2)\right]  
\\&= {N_f L \over 2 \pi} \ln {M \over \Lambda_{PV}} - {2 N_f L \over 2 \pi} \sum\limits_{n>0} \cos({nvL\over 2}) \cos( nuL) (K_0(MLn) - K_0(\Lambda_{PV} L n)),
\end{align}
where we also kept the Pauli-Villars contribution in the last term (here, it can be omitted, but we will need it for later use).
The derivative of the logarithm of the vector supermultiplet determinant, Eq.~(\ref{dets1}), is \cite{Poppitz:2012nz}:
\begin{align}
\label{adj1}
F_{adj} (v,  L)\equiv {d \over d v}  \left[ \log \Delta_{adj} \right]= {6 L \over 2 \pi} \log{\Lambda_{PV} L \over 4 \pi} - {3 L \over 2 \pi}\left[ \psi\left({Lv \over 2 \pi}\right) + \psi\left(1 - {Lv\over 2 \pi}\right)\right],
\end{align}
where $\psi(x) = d \ln \Gamma(x)/dx$.
From (\ref{vkin1},\ref{adj1},\ref{fund1}), we find for the effective 3d coupling:
\begin{align}
\label{3deffective1}
{1 \over g_{3, eff}^2} &= {L \over g_4^2(\Lambda_{PV})} - {L \over 8 \pi^2}(6 - N_f)\log{\Lambda_{PV} L \over 4 \pi}  \nonumber \\
& \; \;+ {3 L \over 8 \pi^2} \left[ \psi\left({Lv \over 2 \pi}\right) + \psi\left(1 - {Lv\over 2 \pi}\right)\right]  \\
& \;  \;+ { N_f L \over  8 \pi^2} \log{4 \pi \over ML} + {2 N_f L \over 8 \pi^2} \sum\limits_{n>0} \cos({nvL\over 2}) \cos( nuL) K_0(MLn)~.\nonumber
\end{align}
The first line  in (\ref{3deffective1}) allows us to replace the bare 4d coupling with its value renormalized at $4 \pi/L$, using the full one-loop beta function of SQCD with $N_f$ flavors ($b_0 = 3N_c - N_f$). The second line represents the contribution of the charged adjoint Kaluza-Klein modes to the effective 3d coupling (or moduli space metric).\footnote{A simple check matching to previous calculations is to study the $L \rightarrow 0$, $v$-fixed, $L/g_4^2(L^{-1})$-fixed limit (i.e., the ``usual" 3d limit). Then, using $\psi(x)\vert_{x\rightarrow 0} \sim - {1 \over x}$, only the contribution of the adjoint Kaluza-Klein zero mode survives, yielding the 3d effective coupling ${L \over g_4^2} - {3  \over 4 \pi v}$, as calculated in, e.g., \cite{Smilga:2004zr}.}
The third line  gives the  contribution of the fundamental multiplets to the moduli space metric. We now study in some more detail its properties. 

\section{Deformation of the moduli-space metric by fundamental flavors}
\label{deformation}

We begin by introducing:
\begin{align}
\label{3deffective3}
\theta &\equiv  v L  \;\; \;\;({\rm separation \; between \; Wilson \; loop \; eigenvalues}) \nonumber \\
m & \equiv M L\;\;  ({\rm dimensionless \; Dirac \; quark \; mass}) \\
\alpha & \equiv u L \; \;\;\;(\alpha=0-{\rm periodic \; quarks}; \alpha=\pi-{\rm antiperiodic \; quarks}) \nonumber\end{align}
We also define $m_{PV} \equiv \Lambda_{PV} L$.
Let us now express
\begin{align}
\label{ffund1}
F_{fund}(v,M,L,u)&={N_f L \over 2 \pi}\left[\log M - 2 \sum\limits_{n>0}\cos({nvL\over2}) \cos(nuL)K_0(LMn) - \left(M \rightarrow \Lambda_{PV}\right)\right] \nonumber \\ 
&= {N_f L \over 2 \pi} \left[\Phi(\theta, m, \alpha) - \Phi(\theta, m_{PV},\alpha)\right]
\end{align}
where
\begin{equation}
\label{phi11}
\Phi(\theta,m,\alpha)  \equiv 
 \log { m}  - 2 \sum\limits_{n>0} \cos({ n \theta \over 2}) \cos({n \alpha}) K_0( m n)~.
\end{equation}
 In what follows, it is useful to rewrite the fundamental contribution in a way that allows to study its small-$M$ and small-$L$ limits. 

We note now various properties of $\Phi(\theta, m, z)$ that may be useful. First, it can be given a representation as a sum over Kaluza-Klein modes, which makes its interpretation as a one-loop correction to the moduli space metric evident. We begin by noting that, from (\ref{ffund1}, \ref{phi11}):
\begin{align}
\label{phi1}
\Phi(\theta,m,\alpha) - \Phi(\theta,m_{PV},\alpha)&= {1 \over 2}  \log {m \over m_{PV}} -   \sum\limits_{n>0} (K_0(m n)-K_0(m_{PV}n)) \cos n ({\theta \over 2} + \alpha)   \nonumber \\
& \;\;+  {1 \over 2}  \log {m \over m_{PV}}  -  \sum\limits_{n>0} (K_0(m n)-K_0(m_{PV}n)) \cos n ({\theta \over 2} -\alpha)~.
\end{align}
Next, we recall from (\ref{fdef},\ref{fexpr}) that:\footnote{It is necessary to take differences,  to cancel infinities in the $s\rightarrow {1\over 2}$ limit and to define the scheme.}
\begin{align}
F({1\over 2}; a, c_1)- F({1\over 2}; a, c_2) =- 2 \log {|c_1|\over |c_2|} + 4 \sum\limits_{n>0}   \cos(2 \pi n a) (K_0(2 \pi n |c_1|)-K_0(2 \pi n |c_2|))~,\nonumber
\end{align}
allowing us to express:
\begin{align}
\label{phi2}
&\Phi(\theta,m,\alpha)- \Phi(\theta,m_{PV},\alpha) \nonumber\\
&=   -{1 \over 4}  F({1\over 2}; {\theta\over 4 \pi} + {\alpha\over 2\pi},{ m\over 2 \pi}) - {1 \over 4} F({1\over 2};  {\theta\over 4 \pi} - {\alpha\over 2\pi},{ m\over 2 \pi})  - (m \rightarrow m_{PV})  \\
& ={1 \over 4} \sum\limits_{k = - \infty}^{\infty} { 1 \over \left[(k + {\theta + 2 \alpha \over 4 \pi})^2 + ({\mu\over 2 \pi})^2 \right]^{1\over 2}} + { 1 \over \left[ (k + {\theta - 2 \alpha \over 4 \pi} )^2 + ({\mu\over 2 \pi})^2 \right]^{1\over 2}} \bigg\vert^{\mu = m_{PV}}_{\mu = m} ,\nonumber
\end{align}
The last equation above  shows that $\Phi$ represents  a one-loop correction to the moduli space metric; a similar representation of the adjoint contribution   was given in \cite{Poppitz:2012nz}.  

To test (\ref{phi2}), we note that it has the correct 4d and 3d limits. The former was already seen in the correct renormalization of the 4d gauge coupling, while to see the latter, we take   $v$ and ${1\over g^2_{3}} = {L\over g_4^2(\Lambda_{PV})}$ fixed, while $L\rightarrow 0$, as well as  $m=0$. Then, we find $\Phi(\theta,0,0) - \Phi(\theta, m_{PV}, 0) = -  {2 \pi \over v}$, which, from (\ref{ffund1}), gives
$F_{fund}(v, 0, L, 0) = -{N_f   \over v}$ and, from (\ref{3deffective1}), a one loop shift ${1\over g^2_{3, eff}} = {1 \over g^2_{3}} + {N_f  \over 4 \pi v}$, which agrees with, e.g., \cite{Smilga:2004zr}. Note that while it would seem that flavors make the gauge coupling weaker as one approaches the origin of moduli space $v \rightarrow 0$, the equation for $g^2_{3, eff} $ shows that quantum corrections overcome the tree level value at $v \le  {4\pi \over N_f g_3^2}$ and hence it should not be trusted beyond this limit; note also that a similar bound for the gaugino supermultiplet contribution applies but with $N_f$ replaced by $3$. Indeed, it is known that the moduli space of even $U(1)$ theories with charged matter is  modified near the origin where Coulomb and Higgs branches meet.

We now come back to the effective 3d coupling (\ref{3deffective1}) and rewrite it as follows (from now on, we denote $g^2 \equiv g_4^2({4 \pi \over L})$):
\begin{align}
\label{3deffective2}
{ 8 \pi^2 \over L g_{3, eff}^2}(\theta,m,\alpha) &= { 8 \pi^2 \over g^2} +   3 \left( \psi({\theta \over 2 \pi})+ \psi(1 - {\theta \over 2\pi}) \right) + N_f  (\log 4 \pi -  \Phi(\theta,m,\alpha))~,
\end{align}
All the $M$-dependence is now in $\Phi$ and we are interested in studying small $m$; as usual, we imagine $L$ is small enough so that $g^2$ is weak and, really, we want $m\ll 1$.
What we are after is to find out, combining (\ref{3deffective2}) with the result from the following section for $\theta_{min} = L v_{min}(m,\alpha,g^2)$ (the supersymmetric expectation value of $v$ for given $M$) and imposing the self-consistency condition that the weak-coupling approximation holds, i.e. that none of the corrections  in (\ref{3deffective2}) is larger than the ``bare" 3d coupling. The latter is naturally identified with the full theory 4d coupling  $\sim {8 \pi^2 \over g^2}$, renormalized at the compactification scale $4 \pi\over L$ and multiplied by $L$. Thus, Eqn.~(\ref{3dweakcondition}) should be useful to give a lower bound on $M$ in each case considered.

\section{Small-$\mathbf M$ asymptotics of ${\mathbf X_{\mathbf{uL}}}$}
\label{next} 
Now we consider the function $X_{uL}(vL,ML)$ and expand around $ML\rightarrow 0$ the function then becomes

\begin{equation}
X_{uL}(vL,ML)=-\frac{2}{\pi}\sum_{n=1}^\infty \frac{1}{n}\sin\left(\frac{nv L}{2}\right)\cos(nuL)[c_1-\ln(ML n)+o(ML)]~,
\end{equation}
where $c_1=-\gamma_E+\ln 2$. The above sum can be rewritten as
\begin{align} \label{xu1}
X_{uL}(vL,ML)&\approx-\frac{1}{\pi}\text{Im}\left\{\sum_{n=1}^\infty \frac{1}{n}\Big[c_1-\ln(ML n))\Big]e^{i n(\frac{v}{2}+u)L}+(u\rightarrow -u)\right\} \\
&= \frac{1}{\pi}\;\text{Im}\left\{\ln(1-e^{i \left(\frac{v}{2}+u\right)L})(c_1-\ln(ML))-\partial_\nu\text{Li}_{\nu+1}(e^{i\left(\frac{v}{2}+u\right)L})\Bigg|_{\nu=0}+(u\rightarrow -u)\right\}~. \nonumber
\end{align}
The above result was obtained from the analytical continuation of the following sums:
\begin{align}
&\sum\limits_{n>0} \frac{1}{n}e^{-x n}=-\ln(1-e^{-x})\;, \nonumber \\
&\sum\limits_{n>0}  \frac{\ln n}{n} e^{-x n}=\sum\limits_{n>0}  \frac{\ln n}{n^{1+\nu}}e^{-n x}\Bigg|_{\nu=0}=-\partial_\nu\sum\limits_{n\ge0}  \frac{1}{n^{1+\nu}}e^{-x n}\Bigg|_{\nu=0}=-\partial_\nu \text{Li}_{1+\nu}(e^{-x})\Bigg|_{\nu=0}~.
\end{align}
Finally, since 
\begin{equation}\label{ab3}
\frac{1}{\pi}\text{Im}\left(\ln(1-e^{i(\frac{v}{2}\pm u)L})\right)=-\frac{1}{2}+\left(\frac{\frac{vL }{2}\pm uL}{2\pi}-\left\lfloor \frac{\frac{vL}{2}\pm uL}{2\pi}\right\rfloor\right)~,
\end{equation}
we obtain that:
\begin{align}
\label{xu3}
X_{uL}(vL,ML)&\approx \left(-1+\frac{vL}{2\pi}-\left\lfloor \frac{\frac{vL}{2}+ uL}{2\pi}\right\rfloor-\left\lfloor \frac{\frac{vL}{2}- uL}{2\pi}\right\rfloor\right)(c_1-\ln(ML))\nonumber \\&\qquad- \frac{1}{\pi}\text{Im}\left\{ \partial_\nu\left[ \text{Li}_{\nu+1}\left(e^{i\left(\frac{vL}{2}+uL\right)}\right)+\text{Li}_{\nu+1}\left(e^{i\left(\frac{vL}{2}-uL\right)}\right)\right]\right\}\bigg\vert_{\nu=0}~.
\end{align}
The logarithmically divergent (as $ML \rightarrow 0$) part of $X$ relevant for SQCD with center symmetry is  consistent with (\ref{centersymmx}):
\begin{align}
\label{xucenterlimit}
X_{{\pi\over 2}+ L \delta u}(\pi, ML)  & +X_{-{\pi\over 2} + L \delta u}(\pi, ML) \\
   \simeq& - \ln(ML)\left(-1 - \left\lfloor {1\over 2} + {\delta u \over 2 \pi} \right\rfloor - \left\lfloor- {\delta u \over 2 \pi} \right\rfloor   - \left\lfloor   {\delta u \over 2 \pi} \right\rfloor - \left\lfloor {1\over 2} - {\delta u \over 2 \pi} \right\rfloor \right)  = 0\nonumber 
\end{align}
for $L \delta u $ between $0$ and $\pi$.
Notice also that the $\log ML$ terms in (\ref{xu3}) combine with the other finite contribution of the fundamental determinant into an expression proportional to $\log LM$ times the index for fundamental fermions in the BPS monopole-instanton background, as shown in Eqs.~(\ref{index1}, \ref{index2}).

\bibliography{FundSYMA}

\providecommand{\href}[2]{#2}\begingroup\raggedright\begin{thebibliography}{10}

\bibitem{Gross:1980br}
D.~J. Gross, R.~D. Pisarski, and L.~G. Yaffe, {\it {QCD and Instantons at
  Finite Temperature}},  {\em Rev.Mod.Phys.} {\bf 53} (1981) 43.

\bibitem{Pisarski:2001pe}
R.~D. Pisarski, {\it {Tests of the Polyakov loops model}},  {\em Nucl.Phys.}
  {\bf A702} (2002) 151--158,
  [\href{http://xxx.lanl.gov/abs/hep-ph/0112037}{{\tt hep-ph/0112037}}].

\bibitem{Fukushima:2003fw}
K.~Fukushima, {\it {Chiral effective model with the Polyakov loop}},  {\em
  Phys.Lett.} {\bf B591} (2004) 277--284,
  [\href{http://xxx.lanl.gov/abs/hep-ph/0310121}{{\tt hep-ph/0310121}}].

\bibitem{Ratti:2005jh}
C.~Ratti, M.~A. Thaler, and W.~Weise, {\it {Phases of QCD: Lattice
  thermodynamics and a field theoretical model}},  {\em Phys.Rev.} {\bf D73}
  (2006) 014019, [\href{http://xxx.lanl.gov/abs/hep-ph/0506234}{{\tt
  hep-ph/0506234}}].

\bibitem{Diakonov:2012dx}
D.~Diakonov, C.~Gattringer, and H.-P. Schadler, {\it {Free energy for
  parameterized Polyakov loops in SU(2) and SU(3) lattice gauge theory}},  {\em
  JHEP} {\bf 1208} (2012) 128, [\href{http://xxx.lanl.gov/abs/1205.4768}{{\tt
  arXiv:1205.4768}}].

\bibitem{Greensite:2012dy}
J.~Greensite, {\it {The potential of the effective Polyakov line action from
  the underlying lattice gauge theory}},  {\em Phys.Rev.} {\bf D86} (2012)
  114507, [\href{http://xxx.lanl.gov/abs/1209.5697}{{\tt arXiv:1209.5697}}].

\bibitem{Haas:2013qwp}
L.~M. Haas, R.~Stiele, J.~Braun, J.~M. Pawlowski, and J.~Schaffner-Bielich,
  {\it {Improved Polyakov-loop potential for effective models from functional
  calculations}},  \href{http://xxx.lanl.gov/abs/1302.1993}{{\tt
  arXiv:1302.1993}}.

\bibitem{Aharony:2006rf}
O.~Aharony, J.~Marsano, and M.~Van~Raamsdonk, {\it {Two loop partition function
  for large N pure Yang-Mills theory on a small $S^3$}},  {\em Phys.Rev.} {\bf
  D74} (2006) 105012, [\href{http://xxx.lanl.gov/abs/hep-th/0608156}{{\tt
  hep-th/0608156}}].

\bibitem{Anber:2012ig}
M.~M. Anber, S.~Collier, and E.~Poppitz, {\it {The $SU(3)/Z_3$ QCD(adj)
  deconfinement transition via the gauge theory/'affine' XY-model duality}},
  {\em JHEP} {\bf 1301} (2013) 126,
  [\href{http://xxx.lanl.gov/abs/1211.2824}{{\tt arXiv:1211.2824}}].

\bibitem{Davies:1999uw}
N.~M. Davies, T.~J. Hollowood, V.~V. Khoze, and M.~P. Mattis, {\it {Gluino
  condensate and magnetic monopoles in supersymmetric gluodynamics}},  {\em
  Nucl.Phys.} {\bf B559} (1999) 123--142,
  [\href{http://xxx.lanl.gov/abs/hep-th/9905015}{{\tt hep-th/9905015}}].

\bibitem{Seiberg:1996nz}
N.~Seiberg and E.~Witten, {\it {Gauge dynamics and compactification to
  three-dimensions}},  \href{http://xxx.lanl.gov/abs/hep-th/9607163}{{\tt
  hep-th/9607163}}.

\bibitem{Aharony:1997bx}
O.~Aharony, A.~Hanany, K.~A. Intriligator, N.~Seiberg, and M.~Strassler, {\it
  {Aspects of N=2 supersymmetric gauge theories in three-dimensions}},  {\em
  Nucl.Phys.} {\bf B499} (1997) 67--99,
  [\href{http://xxx.lanl.gov/abs/hep-th/9703110}{{\tt hep-th/9703110}}].

\bibitem{Unsal:2007jx}
M.~Unsal, {\it {Magnetic bion condensation: A New mechanism of confinement and
  mass gap in four dimensions}},  {\em Phys.Rev.} {\bf D80} (2009) 065001,
  [\href{http://xxx.lanl.gov/abs/0709.3269}{{\tt arXiv:0709.3269}}].

\bibitem{Polyakov:1976fu}
A.~M. Polyakov, {\it {Quark Confinement and Topology of Gauge Groups}},  {\em
  Nucl.Phys.} {\bf B120} (1977) 429--458.

\bibitem{Poppitz:2011wy}
E.~Poppitz and M.~Unsal, {\it {Seiberg-Witten and 'Polyakov-like' magnetic bion
  confinements are continuously connected}},  {\em JHEP} {\bf 1107} (2011) 082,
  [\href{http://xxx.lanl.gov/abs/1105.3969}{{\tt arXiv:1105.3969}}].

\bibitem{Poppitz:2012sw}
E.~Poppitz, T.~Schaefer, and M.~Unsal, {\it {Continuity, Deconfinement, and
  (Super) Yang-Mills Theory}},  {\em JHEP} {\bf 1210} (2012) 115,
  [\href{http://xxx.lanl.gov/abs/1205.0290}{{\tt arXiv:1205.0290}}].

\bibitem{Poppitz:2012nz}
E.~Poppitz, T.~Schaefer, and M.~Unsal, {\it {Universal mechanism of
  (semi-classical) deconfinement and theta-dependence for all simple groups}},
  {\em JHEP} {\bf 1303} (2013) 087,
  [\href{http://xxx.lanl.gov/abs/1212.1238}{{\tt arXiv:1212.1238}}].

\bibitem{Argyres:2012vv}
P.~Argyres and M.~Unsal, {\it {A semiclassical realization of infrared
  renormalons}},  {\em Phys.Rev.Lett.} {\bf 109} (2012) 121601,
  [\href{http://xxx.lanl.gov/abs/1204.1661}{{\tt arXiv:1204.1661}}].

\bibitem{Argyres:2012ka}
P.~C. Argyres and M.~Unsal, {\it {The semi-classical expansion and resurgence
  in gauge theories: new perturbative, instanton, bion, and renormalon
  effects}},  {\em JHEP} {\bf 1208} (2012) 063,
  [\href{http://xxx.lanl.gov/abs/1206.1890}{{\tt arXiv:1206.1890}}].

\bibitem{Dunne:2012ae}
G.~V. Dunne and M.~Unsal, {\it {Resurgence and Trans-series in Quantum Field
  Theory: The CP(N-1) Model}},  {\em JHEP} {\bf 1211} (2012) 170,
  [\href{http://xxx.lanl.gov/abs/1210.2423}{{\tt arXiv:1210.2423}}].

\bibitem{Dunne:2012zk}
G.~V. Dunne and M.~Unsal, {\it {Continuity and Resurgence: towards a continuum
  definition of the CP(N-1) model}},  {\em Phys.Rev.} {\bf D87} (2013) 025015,
  [\href{http://xxx.lanl.gov/abs/1210.3646}{{\tt arXiv:1210.3646}}].

\bibitem{Unsal:2010qh}
M.~Unsal and L.~G. Yaffe, {\it {Large-N volume independence in conformal and
  confining gauge theories}},  {\em JHEP} {\bf 1008} (2010) 030,
  [\href{http://xxx.lanl.gov/abs/1006.2101}{{\tt arXiv:1006.2101}}].

\bibitem{Heller:1984eq}
U.~M. Heller and F.~Karsch, {\it {Finite temperature SU(2) lattice gauge theory
  with dynamical fermions}},  {\em Nucl.Phys.} {\bf B258} (1985) 29.

\bibitem{Lee:1997vp}
K.-M. Lee and P.~Yi, {\it {Monopoles and instantons on partially compactified
  D-branes}},  {\em Phys.Rev.} {\bf D56} (1997) 3711--3717,
  [\href{http://xxx.lanl.gov/abs/hep-th/9702107}{{\tt hep-th/9702107}}].

\bibitem{Kraan:1998pm}
T.~C. Kraan and P.~van Baal, {\it {Periodic instantons with nontrivial
  holonomy}},  {\em Nucl.Phys.} {\bf B533} (1998) 627--659,
  [\href{http://xxx.lanl.gov/abs/hep-th/9805168}{{\tt hep-th/9805168}}].

\bibitem{Anber:2011de}
M.~M. Anber and E.~Poppitz, {\it {Microscopic Structure of Magnetic Bions}},
  {\em JHEP} {\bf 1106} (2011) 136,
  [\href{http://xxx.lanl.gov/abs/1105.0940}{{\tt arXiv:1105.0940}}].

\bibitem{Wess:1992cp}
J.~Wess and J.~Bagger, {\it {Supersymmetry and supergravity, Princeton
  University Press, 1992}}, .

\bibitem{Lucini:2012gg}
B.~Lucini and M.~Panero, {\it {SU(N) gauge theories at large N}},  {\em
  Phys.Rept.} {\bf 526} (2013) 93--163,
  [\href{http://xxx.lanl.gov/abs/1210.4997}{{\tt arXiv:1210.4997}}].

\bibitem{Unsal:2012zj}
M.~Unsal, {\it {Theta dependence, sign problems and topological interference}},
   {\em Phys.Rev.} {\bf D86} (2012) 105012,
  [\href{http://xxx.lanl.gov/abs/1201.6426}{{\tt arXiv:1201.6426}}].

\bibitem{Parnachev:2008fy}
A.~Parnachev and A.~R. Zhitnitsky, {\it {Phase Transitions, theta Behavior and
  Instantons in QCD and its Holographic Model}},  {\em Phys.Rev.} {\bf D78}
  (2008) 125002, [\href{http://xxx.lanl.gov/abs/0806.1736}{{\tt
  arXiv:0806.1736}}].

\bibitem{Thomas:2011ee}
E.~Thomas and A.~R. Zhitnitsky, {\it {Topological Susceptibility and Contact
  Term in QCD. A Toy Model}},  {\em Phys.Rev.} {\bf D85} (2012) 044039,
  [\href{http://xxx.lanl.gov/abs/1109.2608}{{\tt arXiv:1109.2608}}].

\bibitem{Anber:2013sga}
M.~M. Anber, {\it {Theta dependence of the deconfining phase transition in pure
  SU(Nc) Yang-Mills theories}},  \href{http://xxx.lanl.gov/abs/1302.2641}{{\tt
  arXiv:1302.2641}}.

\bibitem{D'Elia:2012vv}
M.~D'Elia and F.~Negro, {\it {$\theta$ dependence of the deconfinement
  temperature in Yang-Mills theories}},  {\em Phys.Rev.Lett.} {\bf 109} (2012)
  072001, [\href{http://xxx.lanl.gov/abs/1205.0538}{{\tt arXiv:1205.0538}}].

\bibitem{D'Elia:2013eua}
M.~D'Elia and F.~Negro, {\it {On the phase diagram of Yang-Mills theories in
  the presence of a theta therm}},
  \href{http://xxx.lanl.gov/abs/1306.2919}{{\tt arXiv:1306.2919}}.

\bibitem{Shuryak:2013tka}
E.~Shuryak and T.~Sulejmanpasic, {\it {Holonomy potential and confinement from
  a simple model of the gauge topology}},
  \href{http://xxx.lanl.gov/abs/1305.0796}{{\tt arXiv:1305.0796}}.

\bibitem{Pepe:2006er}
M.~Pepe and U.-J. Wiese, {\it {Exceptional Deconfinement in G(2) Gauge
  Theory}},  {\em Nucl.Phys.} {\bf B768} (2007) 21--37,
  [\href{http://xxx.lanl.gov/abs/hep-lat/0610076}{{\tt hep-lat/0610076}}].

\bibitem{Cossu:2007dk}
G.~Cossu, M.~D'Elia, A.~Di~Giacomo, B.~Lucini, and C.~Pica, {\it {G(2) gauge
  theory at finite temperature}},  {\em JHEP} {\bf 0710} (2007) 100,
  [\href{http://xxx.lanl.gov/abs/0709.0669}{{\tt arXiv:0709.0669}}].

\bibitem{Chen:2004tb}
H.-S. Chen and X.-Q. Luo, {\it {Phase diagram of QCD at finite temperature and
  chemical potential from lattice simulations with dynamical Wilson quarks}},
  {\em Phys.Rev.} {\bf D72} (2005) 034504,
  [\href{http://xxx.lanl.gov/abs/hep-lat/0411023}{{\tt hep-lat/0411023}}].

\bibitem{Poppitz:2008hr}
E.~Poppitz and M.~Unsal, {\it {Index theorem for topological excitations on
  $R^3 \times S^1$ and Chern-Simons theory}},  {\em JHEP} {\bf 0903} (2009)
  027, [\href{http://xxx.lanl.gov/abs/0812.2085}{{\tt arXiv:0812.2085}}].

\bibitem{Aharony:2013dha}
O.~Aharony, S.~S. Razamat, N.~Seiberg, and B.~Willett, {\it {3d dualities from
  4d dualities}},  \href{http://xxx.lanl.gov/abs/1305.3924}{{\tt
  arXiv:1305.3924}}.

\bibitem{Beasley:2004ys}
C.~Beasley and E.~Witten, {\it {New instanton effects in supersymmetric QCD}},
  {\em JHEP} {\bf 0501} (2005) 056,
  [\href{http://xxx.lanl.gov/abs/hep-th/0409149}{{\tt hep-th/0409149}}].

\bibitem{Davies:2000nw}
N.~M. Davies, T.~J. Hollowood, and V.~V. Khoze, {\it {Monopoles, affine
  algebras and the gluino condensate}},  {\em J.Math.Phys.} {\bf 44} (2003)
  3640--3656, [\href{http://xxx.lanl.gov/abs/hep-th/0006011}{{\tt
  hep-th/0006011}}].

\bibitem{Intriligator:2013lca}
K.~Intriligator and N.~Seiberg, {\it {Aspects of 3d N=2 Chern-Simons-Matter
  Theories}},  \href{http://xxx.lanl.gov/abs/1305.1633}{{\tt arXiv:1305.1633}}.

\bibitem{Seiberg:1994aj}
N.~Seiberg and E.~Witten, {\it {Monopoles, duality and chiral symmetry breaking
  in N=2 supersymmetric QCD}},  {\em Nucl.Phys.} {\bf B431} (1994) 484--550,
  [\href{http://xxx.lanl.gov/abs/hep-th/9408099}{{\tt hep-th/9408099}}].

\bibitem{Yung:1987zp}
A.~Yung, {\it {Instanton Vacuum in Supersymmetric QCD}},  {\em Nucl.Phys.} {\bf
  B297} (1988) 47.

\bibitem{Schafer:1996wv}
T.~Schaefer and E.~V. Shuryak, {\it {Instantons in QCD}},  {\em Rev.Mod.Phys.}
  {\bf 70} (1998) 323--426, [\href{http://xxx.lanl.gov/abs/hep-ph/9610451}{{\tt
  hep-ph/9610451}}].

\bibitem{Shuryak:2012aa}
E.~Shuryak and T.~Sulejmanpasic, {\it {The Chiral Symmetry Breaking/Restoration
  in Dyonic Vacuum}},  {\em Phys.Rev.} {\bf D86} (2012) 036001,
  [\href{http://xxx.lanl.gov/abs/1201.5624}{{\tt arXiv:1201.5624}}].

\bibitem{Faccioli:2013ja}
P.~Faccioli and E.~Shuryak, {\it {QCD Topology at Finite Temperature:
  Statistical Mechanics of Selfdual Dyons}},
  \href{http://xxx.lanl.gov/abs/1301.2523}{{\tt arXiv:1301.2523}}.

\bibitem{Vandoren:2008xg}
S.~Vandoren and P.~van Nieuwenhuizen, {\it {Lectures on instantons}},
  \href{http://xxx.lanl.gov/abs/0802.1862}{{\tt arXiv:0802.1862}}.

\bibitem{Ponton:2001hq}
E.~Ponton and E.~Poppitz, {\it {Casimir energy and radius stabilization in
  five-dimensional orbifolds and six-dimensional orbifolds}},  {\em JHEP} {\bf
  0106} (2001) 019, [\href{http://xxx.lanl.gov/abs/hep-ph/0105021}{{\tt
  hep-ph/0105021}}].

\bibitem{Smilga:2004zr}
A.~V. Smilga and A.~Vainshtein, {\it {Background field calculations and
  nonrenormalization theorems in 4-D supersymmetric gauge theories and their
  low-dimensional descendants}},  {\em Nucl.Phys.} {\bf B704} (2005) 445--474,
  [\href{http://xxx.lanl.gov/abs/hep-th/0405142}{{\tt hep-th/0405142}}].

\end{thebibliography}\endgroup
\bibliographystyle{JHEP}

\end{document}